\documentclass[11pt,preprint]{aastex}

\newcommand{\masr}{mas~yr$^{-1}$}

\slugcomment{to appear in The Astrophysical Journal}

\shortauthors{Hardee, Walker, \& G\'omez}
\shorttitle{Modeling the 3C\,120 jet}

\begin{document}

\input colordvi.sty
 
\baselineskip 12pt
\parskip 0pt


\title{MODELING THE 3C\,120 RADIO JET FROM 1 TO 30 MilliARCSECONDS}

\author{P.E. Hardee}
\affil{Department of Physics \& Astronomy, The University of Alabama,
Tuscaloosa, AL 35487, USA}
\email{hardee@athena.astr.ua.edu}

\author{R.C. Walker}
\affil{National Radio Astronomy Observatory\footnote{The National
Radio Astronomy Observatory is a facility of the National Science
Foundation, operated under cooperative agreement by Associated
Universities, Inc.}, Socorro, NM 87801, USA}
\email{cwalker@nrao.edu}

\author{J. L. G\'omez}
\affil{Instituto de Astrof\'{\i}sica de Andaluc\'{\i}a (CSIC), Apartado 3004, Granda 18080, Spain,  \and Institut d'Estudis Espacials
de Catalunya/CSIC, Edif. Nexus, Gran Capita 2-4, 08034 Barcelona, Spain}
\email{jlgomez@iaa.es}


\begin{abstract}
\vspace{-0.4cm}
\baselineskip 12pt
\parskip 0pt

In this paper we use the predicted spatial development of helical
structures along an expanding jet to model observed structures and
motions in the 3C\,120 jet.  New results of VLBI imaging of the
parsec-scale radio jet in 3C\,120 at 5 GHz are examined along with
older long term monitoring results at 5 GHz and older results obtained
at 22 and 43 GHz.  The high frequency observations provide detailed
information on motions and structure from 0.5 to 10~mas from the core
and the lower frequency observations from 1 to 30~mas from the core.
Proper motions of helical components associated with the pattern and of
other components that move through the pattern provide estimates of
flow and helical pattern speeds.  Theoretical modeling of the motion
and appearance of the helical pattern allows determination of sound
speeds as a function of the jet viewing angle.  The jet sound speed
declines although probably not as fast as adiabatically.  At a
12\arcdeg\ viewing angle the most likely scenario involves a decline in
jet sound speed from $c/3 < a_j < c/\sqrt 3$ at $\sim 0.5$~mas from the
core to $0.1 c < a_j < 0.25c$ at $\sim 25$~mas from the core
accompanied by some acceleration in the jet flow from Lorentz factor
$\gamma \lesssim 5$ to $\gamma \gtrsim 7$.  The sound speed in the
cocoon medium around the jet is less well determined but is less than
the sound speed in the jet probably by a factor of 1.5 - 5.  A largest
possible viewing angle of 15\arcdeg\ implies a jet sound speed at the
upper limit of these estimates and somewhat higher flow Lorentz
factors. However, jet morphology argues against viewing angles larger than
12\arcdeg. At smaller viewing angles the jet sound speed is lower and at
a 6\arcdeg\ viewing angle the jet sound speed is about a factor 2 less
but the flow Lorentz factor is comparable. The decline in radio
intensity is on the order of what would be associated with isothermal
jet expansion.  Knot interknot intensity variations are greater than
would be expected from adiabatic compressions associated with the
helical twist and we infer the presence of a shock along the leading
edge of the helical twist in addition to shock or density structures
flowing through the helical pattern. Our results imply that the
macroscopic heating of the expanding jet fluid is less than the
microscopic energization of the synchrotron radiating relativistic
electrons.

\end{abstract}

\keywords{galaxies: individual (3C\,120) --- galaxies: jets --- 
galaxies: active --- radio continuum: galaxies --- hydrodynamics --- 
relativity \vspace{-0.5cm}}

\section{Introduction}
\baselineskip 12pt
\parskip 2pt

The radio source 3C\,120 is dominated by a variable core and a
prominent one-sided jet extending from the core on subparsec scales to
about a hundred kiloparsecs \citep{W87}.  The galaxy has a redshift
$z=0.033$ \citep{B80}, and with $H_o = h~ 100$~km~s$^{-1}$~Mpc$^{-1}$
and $h/0.7 \equiv h_{70} = 1$ there are $\sim 0.65$~pc~mas$^{-1}$.  Recent high
frequency VLBI observations at 86, 43 and 22 GHz
\citep{G98,G99,G00,G01,M02} have produced high resolution images of the
innermost features and motions in the jet.   The 43 and 22 GHz
observations, show that the inner jet structure, within 10~mas from the
core, is complex and exhibits a subluminal to superluminal transition
on very small scales. Global VLBI observations at 5~GHz and 10.7~GHz
\citep{W01} find superluminal motions consistent with
those observed at 22 and 43~GHz.  The 3C\,120 jet shows a variety of
knots and side-to-side structures, and it has been suggested that the
jet structure of 3C\,120 inside 10~mas may be described by components
interacting with the external medium in their motion along an
underlying helical twist \citep{G98,G00,G01}. VLBI images at 1.7~GHz,
see Figure 6 in \citet{W01}, also show structure suggestive of a
helical twist beyond 25~mas but with considerably longer wavelength.

Helical structure in relativistic jets can arise as a result of ordered
variation in the flow direction at the central engine, e.g.,
precession, and/or as a result of random perturbations to the jet flow
such as the jet cloud interaction discussed in \citet{G00}.  Initial
random perturbations can trigger or propagate as pinch, helical, or
higher order normal modes of jet distortion.  Any normal mode structure
will depend on the initial excited wave frequency or frequencies and
initial amplitudes, and on the subsequent propagation and growth or
damping of these wave frequencies along the jet, e.g., axisymmetric
perturbations as in \citet{H98} and \citet{A01} or some combination
including asymmetric perturbations as in \citet{A03} and \citet{HH03}.
In these simulations linear and/or non-linear perturbations were
applied.  In particular, an injection event can coexist and interact
with a precession induced helical twist \citep{A03}. In general,
non-linearly triggered normal modes are damped until reaching a
``linear'' regime and both non-linear triggers and ``linear'' normal
mode structures can coexist on the jets. Within the ``linear'' regime
normal mode structures can be satisfactorily modeled by the linearized
relativistic hydrodynamic (RHD) equations \citep{H98,A01,HH03}.   The
behavior of normal modes and, in particular, helical twist propagation
and growth or damping can sensitively depend on jet speed via the
Lorentz factor, on the sound speeds in the jet and surrounding
material, and on the rate of jet expansion \citep{H03}.

That the normal wave modes predicted theoretically operate on AGN jets
has been suggested by \citet{LZ01} who successfully fitted the
twisted emission threads observed in the 3C\,273 jet by a combination
of helical and elliptical surface and internal normal modes.  Other
similar fitting has been performed by \citet{LHE03} in the context of
M\,87.  Helical twists have been invoked to explain other observed jet
structure and, for example, it has been shown that superluminal motions
and accelerations along a curved trajectory can be produced by helical
jet models, e.g., the radio source 3C\,345 \citep{H87,S95}.  While some
fitting of helically twisted structure to observed jet morphology has
been performed, to date no detailed self-consistent models have been
constructed.

Only in the past few years have time dependent relativistic
hydrodynamical codes become readily available
\citep{DH94,MMI94,KNM96,FK96,A99a}.  More recently this has allowed
fully 3D relativistic jet simulations to be performed with resolution
sufficient to compare structure with theoretical predictions and/or
with features observed in AGN jets
\citep{A99b,A00,H01,HMD02,HH03,A03}.  Ideally we would like to model
observed jet structures numerically and include non-linear effects. In
general, computational constraints make it very difficult to simulate
and model the superluminal jets that must lie near to the line of
sight.  Satisfactory numerical modeling would require very high grid
resolution across the jet to reduce the effects of numerical viscosity
($\sim 100$ computational zones) \citet{P04} and additional zones
outside the jet, lengthy computational grids in order to account for
projection, and very large storage capabilities in order to account for
light travel time effects, e.g., \citet{G97,A03}, that make it
impractical to conduct numerical modeling involving many trials.  This
difficulty can be overcome using theoretical models based on the
linearized fluid equations.

In this paper we consider high frequency VLBI images and proper motions
from \citet{G98,G99,G01} that cover the inner 10~mas of the jet, new
5~GHz images that cover the inner jet out to 30~mas, and proper motions
from \citet{W01} that cover this portion of the jet.   With these
observations, we begin to discern what trajectory components follow,
whether motions are constant for a given feature and between features,
and whether there are similarities between component speeds,
trajectories, and structure at parsec and tens of parsec scales.  We
use the observed structure and proper motions to constrain the
macroscopic properties of the jet and surrounding medium.  In \S 2 the
observed relevant observations are summarized.  In \S 3 we show how to
model relativistically moving helical twists along with the
accompanying line of sight appearance and model the innermost 10~mas
portion of the jet as a single frequency helical twist on an
isothermally expanding jet.  In \S 4 we extend the modeling to larger
distances with dual harmonic twist frequencies and also consider
adiabatic expansion.  We conclude in \S 5 with a summary and discussion
of the implications of our intensity and dynamical modeling.

\vspace{-0.7cm}
\section{Observed Structure and Motion inside 30~mas}
\vspace{-0.1cm}
  
The inner 10~mas structure of 3C\,120 has been interpreted as
determined by the evolution of superluminal components and their
complex interactions with the external medium and/or underlying jet
\citep{G98,G99,G00,G01,M02} The association of some of these components
with injection of material into the jet, leading to the formation of
shocks, has been established by simultaneous radio and X-ray
observations. Dips in the X-ray light curve are interpreted as caused
by the disappearance of a section of the inner accretion disk past the
event horizon of the black hole, while the remainder of the disk
material is injected into the jet, leading to a flare in the radio
light curves and the appearance of a new superluminal component in the
jet \citep{M02}.  Other components seem to be generated by the passage
down the jet of such leading shocks, as has been considered for
components $m$, $r$, and $s$ \citep{G01}. These are observed to appear
in the wake of the strong superluminal component $o$, and present a
similar behavior to the {\it trailing} components obtained in numerical
simulations by \citet{A01}.

Continued monthly monitoring of the inner structure at 22 and 43 GHz
with the VLBA has revealed rapid changes in the total and linearly
polarized intensity, accompanied by a rotation of the magnetic
polarization vector, interpreted as resulting from the interaction of
the jet components with the external medium \citep{G00,G01}. In
particular, these high frequency observations also show evidence for
the existence of an underlying helical jet structure, which may also be
affecting the jet's structural evolution. Component $o$ (Figure 1 at 22
GHz) can be resolved into two subcomponents $o1$ and $o2$ (Figure 1 at
43 GHz) that show changes in the position angle and magnetic
polarization suggesting interaction with an underlying helical twist
\citep{G01}.
\begin{figure}[h]
\vspace{8.9cm}
\includegraphics{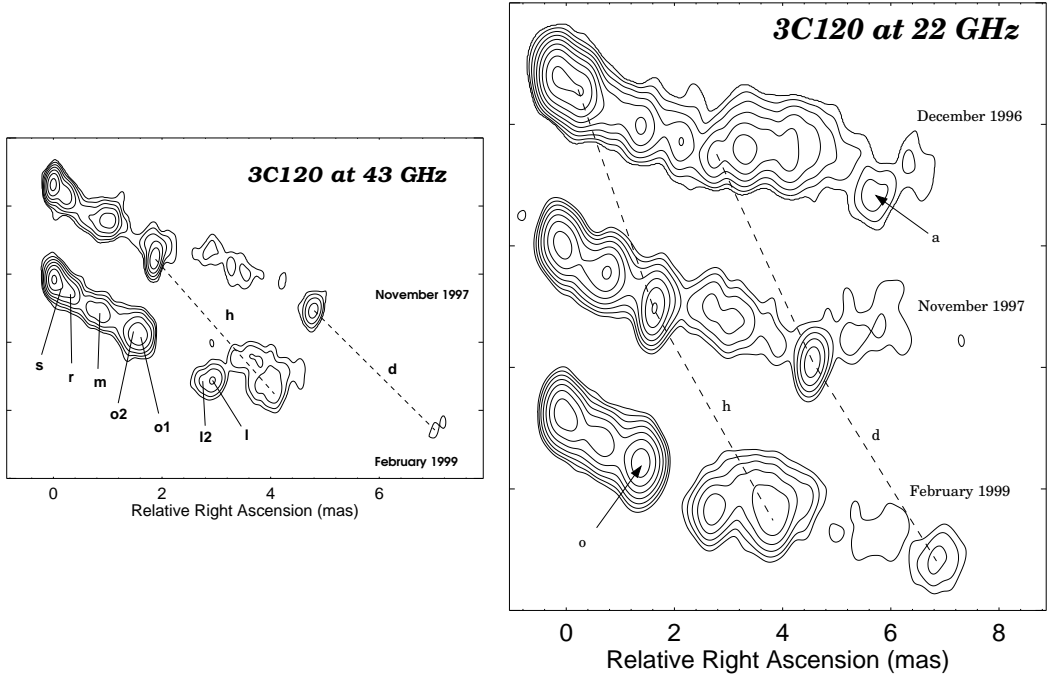}
\caption{\footnotesize \baselineskip 10pt Multiepoch VLBI images of 3C\,120 at 43 and 22 GHz with the
various components labeled. At 43~GHz the contour levels start at
4.4~mJy beam$^{-1}$ with increase by factors of 2 from there. The
intensity peaks at 0.37 Jy~beam$^{-1}$ (Nov 1997) and 0.33
Jy~beam$^{-1}$ (Feb 1999).  The convolving beam is $0.35 \times
0.16$~mas extended in position angle near 9 degrees.  At 22~GHz the
contour levels start at 4.2~mJy beam$^{-1}$ except for Dec 1996 which
starts at 2.1~mJy beam$^{-1}$ and increase by factors of 2 from there.
The intensity peaks at  0.67 Jy~beam$^{-1}$ (Dec 1996), 0.38
Jy~beam$^{-1}$ (Nov 1997), and 0.43 Jy~beam$^{-1}$ (Feb 1999).  The
convolving beam is $0.6 \times 0.3$~mas extended in position angle near
9 degrees.
\label {fig1}}
\vspace{-0.2cm}
\end{figure}
In general, there are "northern" and "southern" jet components, see
Figures 2 in \citet{G00,G01}, that seem to be ejected with
different position angles.  The southern components are usually
brighter than the northern components and this could be the result of the overall curvature of the jet to the north, e.g., \citet{W01},
leading to higher pressures along the southern edge combined with
projection effects associated with a helical twist of the jet.

It is tempting to identify prominent southern components as
indicating an overall helical twist.  The spacing
between the prominent southern components $h,~d$, and $a$ (22~GHz) is
similar and other components can appear to outline a projected helical
twist between these prominent southern components.  For example, note
the curved path in intensity contours between components $d$ and $a$
(December 1996) and between components $h$ and $d$ (November 1997).
That prominent components always evolve towards the southern side of
the jet suggests that projection plays a significant role in their
location.  These ``helical'' components with similar spacing and motion
would indicate the wavelength and motion of the helical twist.  Other
non-helical components would outline and indicate flow through the
helically twisted jet.

Components $h,~d$, and $a$ have separations of $2.5 - 3.5$~mas at core
distances from $2 - 8$~mas with a suggestion of increasing separation
as core distance increases.  Observed proper motions are ($h$) 1.75 and
($d$) 1.71~\masr\ \citep{G99,G01}, and we use an average proper motion
of $\sim$~1.73~\masr\ for these components as representative of motion
of the helical twist.  The relatively constant spacing and motion along
with a ``saturated'' component intensity structure beyond 2~mas
suggests that the motion and spacing of these components with
$\beta_w^{ob} \lesssim 4$, where 1~\masr\ $= 2.2
h_{70}^{-1}$ c, should be associated with a saturated helical twist in
the high frequency regime \citep{H03}.  Here high frequency is relative
to the fastest growing or ``resonant'' frequency of the
Kelvin-Helmholtz unstable helical twist.

Non-helical components can move faster than the helical components.
For example,  in \citet{G01} fast moving components between $o$ and $h$
(e.g., $l$ and $l2$ in Figure 1) show motions from 2.17 to 2.48~\masr.
A wide variety of motions, from $\sim 1.5 - 3.5$~\masr\, are found by
\citet{W01} at 5 and 10.7~GHz out to 14~mas from the core and are shown
along with estimated errors in Figure 2.  An average motion associated
with fast moving components might be $\sim 2.7$~\masr.  Here this
estimate lies between the 2.17~\masr\ from \citet{G01,W01} and the
3.47~\masr\ from \citet{W01} and represents a simple proper motion
average of components B to J in Figure 2. The resulting apparent
superluminal motion, $\beta_f^{ob} \sim 6$, might be assumed to
correspond to the flow speed and constrains the viewing angle to less
than $19 \arcdeg$.
\begin{figure}[h]
\vspace{8.0cm}
\includegraphics{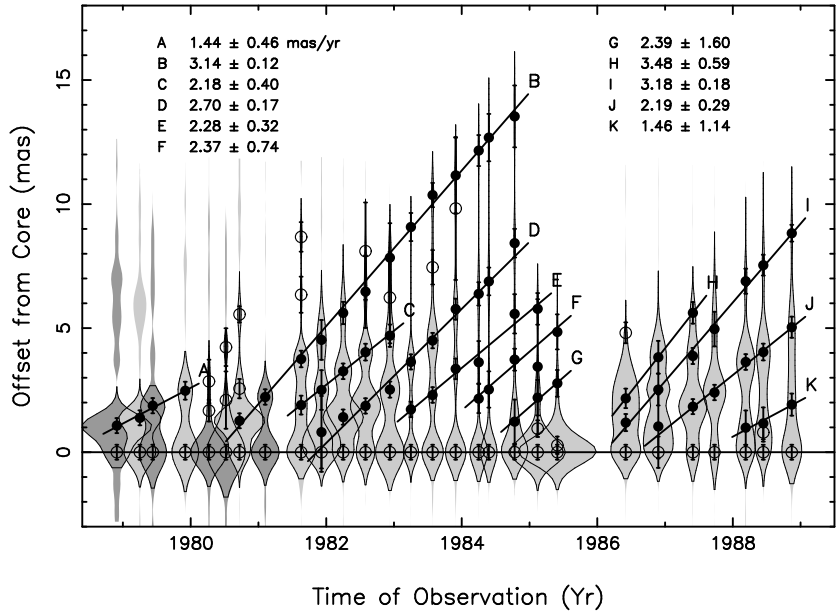}
\caption{\footnotesize \baselineskip 10pt
Motions of components seen at 5 and 10.7~GHz in 3C\,120 between 1978 and 1988.  The components are labeled and a straight line with slopes corresponding to the fitted proper motion is drawn through the points. The proper motion for each feature along with formal errors is written on the figure.  Component labels here do not correspond to components in Figs.\ 1 \& 3.  For additional details see the Figure 4 caption and Table 3 in Walker et al.\ (2001). 
\label {fig2}}
\vspace{-0.3cm}
\end{figure}

Interestingly, component $o$ has a proper motion, 1.83~\masr, similar
to prominent southern components, and spacing between $o$ and $h$ is
similar to the spacing between $h,~d$, and $a$.  Component $o$ appeared
in 1998 after a strong outburst in December 1997 \citep{G01,M02}. Thus,
component $o$ should not necessarily be related to the helical
components $h,~d$, and $a$ assumed formed by projection of the helical
pattern. Also component $o$ is not clearly on the southern edge of the
jet.  Thus, we do not now identify $o$ as a helical component. It will
be interesting to see if in future it behaves like $h,~d$, and $a$.
Such a scenario might occur if a non-uniform outflow is precessed and
components trigger/merge with the overall helical twist.  In this case
component shocks can lead to a shock along the leading edge of the
helical twist.

Components $s, r$, and $m$, interpreted as structures triggered by passage
down the jet of a shock associated with component $o$ \citep{G01}, have
motions $\approx$~0.27, 0.40, and 0.49~\masr\, respectively, and the
separation of these components increases from $s$ to $o$.  Acceleration
and increasing spacing occurs naturally for pinch body mode components
triggered near to the resonant frequency as the pinches transition from
the resonant to the high frequency regime by virtue of jet expansion,
e.g., \citet{A01}.  Alternatively these components could be associated
with the helical twist.  Here again acceleration and increasing spacing
results from a change in helical wave speed as the twist transitions
from the low or resonant frequency regime to the high frequency regime
by virtue of jet expansion \citep{H03}.  In either case the proper
motion of the component nearest to the core with $\beta_w^{ob}
\sim$~0.6 could represent motion of pinching or helical twisting in the
resonant or low frequency regime.  We note here, and address this issue
in \S 3.1, that the proper motion of pinch body mode components or
``helical '' components  associated with the helical surface (twist)
mode are similar in the resonant and high frequency regimes.  Thus, the
identification of these innermost components either with pinches or
helical twist can give a similar final modeling result.

\begin{figure}[h]
\vspace{9.0cm}
\includegraphics{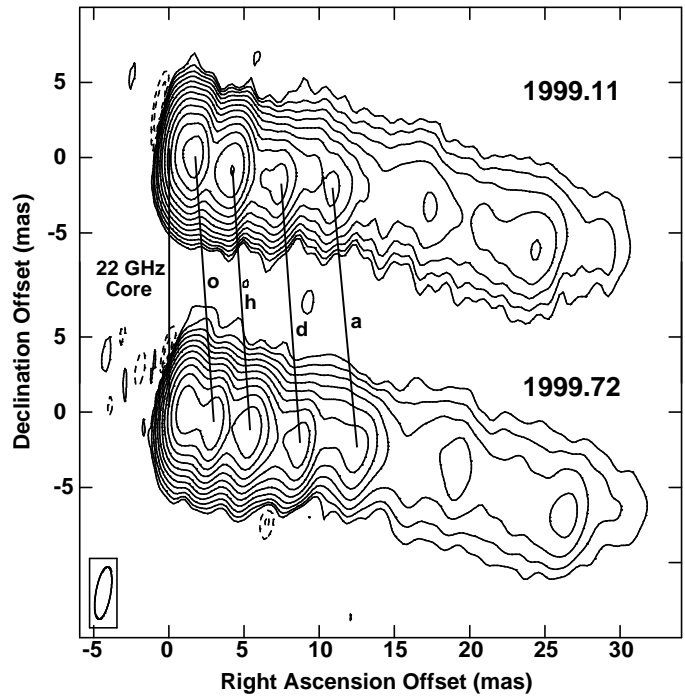}
\caption{\footnotesize \baselineskip 10pt VLBI images of 3C~120 at 5 GHz from 1999.11 and 1999.72 September 21.
The contour levels start with -1, -0.5, 0.5, and 1~mJy
beam$^{-1}$ with increase by factors of 2 from there.  The
peak flux densities are 1.59 and 1.00 Jy beam$^{-1}$ in the first and
second images respectively.  The convolving beams are $3.5 \times 1.0$
mas extended in position angle near 9 degrees.  The marked jet
features in the 1999.11 image correspond to features seen in the
1999 February image of Figure~1, which is based on 22 GHz data from
the preceeding day.  The core position marked here is at the position
of the bright, easternmost feature of the 22 GHz image.  Clearly the
core is significantly absorbed at 5 GHz.
\label {fig3}}
\vspace{-0.4cm}
\end{figure}

Inspection of images in \citet{G00,G01} shows that the jet is bent to
the north inside 10~mas with a position angle change on the order of
$\Delta_{PA}^{ob} \sim 12\arcdeg$, although most of this change appears
to occur beyond 2~mas from the core.  At an assumed viewing angle of
$\theta \sim 10\arcdeg$, this apparent change coresponds to a intrinsic
change of $\Delta_{PA} \sim \sin\theta~\Delta_{PA}^{ob} \sim
2\arcdeg$.  The apparent and intrisic bending angles corresponding to
this position angle change are somewhat larger, i.e., the change in
angle, $\Delta$, defined by a vector tangent to the jet is greater than
the change in position angle, $\Delta_{PA}$, defined by the vector from
the core to a point on the curved jet.  An intrinsic bending angle of
say $\Delta = 3\arcdeg$, even if all bending were to occur in the
innermost 2~mas could not account for the observed acceleration.
Provided $\Delta << \theta << 1$~radian, the change in motion
$\delta\beta^{ob}/\beta^{ob} \sim \Delta/\theta << 1$ and jet bending
to the north cannot account for the observed acceleration from
$\beta^{ob} \sim 0.6$, component $s$, to $\beta^{ob} \sim 1.1$,
component $m$, and falls far short of an acceleration to $\beta^{ob}
\sim 4$, component $o$.

Jet structure out to $\sim 30$~mas is revealed by the total intensity
maps in Figure 3.  The observational details for these previously
unpublished images are given in Appendix A.  Here we identify the first
four components with $o,~h,~d$, and $a$.  In Figure 3 intensity
contours between $h,~d$, and $a$ indicate a helical morphology like
that suggested by \citet{G98} and again with a suggestion of increasing
spacing (typical spacing $\approx 3.5$~mas) as core distance
increases.  Between 10 and 15~mas there is a discontinuity in the
regular pattern of components.  If we interpret the outermost structure
in terms of a helical pattern then the wavelength has shown an abrupt
increase to $\sim 8$~mas.  This increase in helical component spacing
is also accompanied by an apparent increase in the proper motions of
components where component motions beyond about 15~mas from the core
determined from 1.7~GHz multiepoch images are between 2.5 and
3.1~\masr\ \citep{W01}.  This would imply an increase in observed
helical wave speed from $\beta_w^{ob} \lesssim 4$ to $\beta_w^{ob} \sim
6$.

A 1.7~GHz image shown in Figure 4 shows structure between 30 - 80~mas
that is qualitatively similar in appearance to the helical morphology
inside 30~mas. Note the prominent components are again on the southern
side of the jet and note a curved path in intensity contours between
50 - 80~mas from the core. %
\begin{figure}[h]
\vspace{5.0cm}
\includegraphics{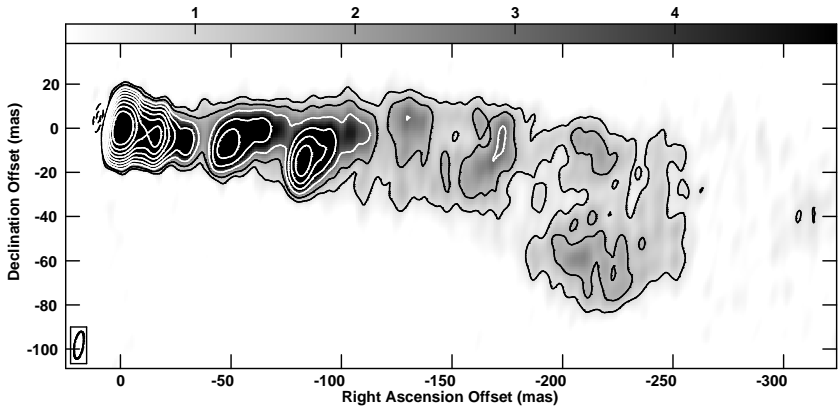}
\caption{\footnotesize \baselineskip 10pt A 1997.70 image of the inner 360~mas of the jet at 1.7~GHz. The image has been convolved to a beam of $12.5~{\rm mas} \times 4.0~{\rm mas}$ elongated in position angle -10\arcdeg. The lower contour levels are -1.5, -0.75, 0.75, and 1.5~mJy~beam$^{-1}$ with the higher levels increasing by factors of 2 from there. 
\label {fig4}}
\vspace{-0.1cm}
\end{figure}
Here we see an indication of a component
spacing jump from $\sim$~15~mas between 35 - 50~mas from the core to
$\sim$~30~mas between 50 - 80~mas from the core.  Beyond the slowly
moving component, $\beta^{ob} < 1$, at $\sim$~80~mas, the jet character
changes significantly. This image and unpublished 5~GHz images suggest
spacing jumps by factors of two as core distance increases.   At least
approximately helical component spacing makes a transition from an
apparent $\lesssim$~3.5~mas spacing within 10~mas from the core to
$\sim$~30~mas by 80~mas from the core.  The implied increase in helical
component spacing, approximately proportional to the jet radius, would
require a systematic linear increase in flow speed as a function of
core distance if it was a flow speed effect.  On the other hand, this
increasing component spacing may be representative of excitation of
multiple harmonic helical frequencies and with lower frequencies
supplanting higher frequencies at larger core distance.

It is from the analysis of the wavelength and motion of helical
structure along with the motion associated with the jet flow that we
expect to constrain, through our modeling, the macroscopic properties
of the jet and surrounding medium.

\vspace{-0.5cm}
\section{Modeling Helical Structures}

Comparison between theoretical predictions and numerical simulations
indicates that helical structures must be operating near to the linear
regime if jets are to remain highly collimated.  In the non-linear
regime helical jet surface distortion at resonant or longer wavelengths
leads to mass entrainment that can disrupt highly collimated flow and
can lead to the formation of an extended velocity shear layer.  Note
however, that helical surface distortion at wavelengths much shorter
than resonant may be able to reach a non-linear saturation without
disrupting highly collimated flow, e.g., \citet{XHS00}. Even when
formally weakly non-linear, normal mode structures can be satisfactoriy
modeled using the linearized fluid equations \citep{H98,HH03}.  In the
linear or weakly non-linear regime moving helical structures and
accompanying flow fields found from the linearized RHD equations behave
in predictable ways and the resulting behavior can lead to estimates of
sound speeds from observed proper motions.  In what follows we show in
detail how to use the observed proper motions of features in the inner
30~mas of the 3C\,120 jet to obtain the first order consistent flow
field associated with the helical structure, produce pseudo-synchrotron
intensity images appropriate to adiabatic compressions associated with
helical structure, and most importantly constrain the parameter regime
in which the jet must operate.

\vspace{-0.7cm}
\subsection{Moving Helical Patterns}
\vspace{-0.1cm}

The linearized RHD equations show that a helical twist is
Kelvin-Helmholtz unstable and propagates at a wave speed dependent on
jet speed, sound speeds, and on the frequency of the wave relative to a
``resonant'' or maximally unstable frequency, $\omega^*$.  For
sufficiently supersonic flow, i.e., $M_j \equiv v_j/a_j >> 1$ and  $M_x
\equiv v_j/a_x >> 1$, the resonant frequency is $\omega^* \sim 1.5
a_x/R_j$ where $R_j$ is the jet radius and $a_{j,x}$ is the sound speed
in jet or external medium.  In general, solution of the wave dispersion
relation is necessary to compute the wave speed and wavelength when the
wave frequency is within an order of magnitude of $\omega^*$.  Along an
expanding jet $\omega^* \propto R_j^{-1}$ decreases and
solution of the dispersion relation is needed to follow the development
of a helical wave of constant frequency as it is advected along the jet
from the low to the high frequency regime.

At frequencies more than an order of magnitude below $\omega^*$
($\omega << \omega^*$) the wave speed is given by 
\begin{equation}
v_w \equiv (\omega/k)\vert_{Real} \approx {\gamma^2 \eta \over 1+\gamma^2 \eta }v_j~,
\end{equation}
and at high frequencies when $\omega >> \omega^*$ the wave speed is given by
\begin{equation}
v_w \equiv (\omega/k)\vert_{Real} \approx \frac{v_j - a_{j}}{1 -
a_{j}v_j/c^2}~. 
\end{equation}
In equations (1) \& (2) $\gamma = (1 - v_j^2/c^2)^{-1/2}$ is the
Lorentz factor, $\eta \equiv (a_{x}/a_{j})^2$, and the sound speed
$a\equiv [ \Gamma P_0 / (\rho _0+[\Gamma /(\Gamma-1 )]P_0/c^2) ]
^{1/2}$, where $4/3\leq \Gamma \leq 5/3$ is the adiabatic index.  The
density, $\rho_0$, and pressure, $P_0$, are measured in the proper
fluid frames, and since pressure balance has been assumed $\eta$ is an
enthalpy ratio if the adiabatic indices are the same inside and outside
the jet.

The apparent helical wavelength in the observer frame is related to the
intrinsic wavelength by $\lambda^{ob}=(\beta^{ob}_w/\beta_w) \lambda$ where
$\beta^{ob}_w \equiv [\sin\theta/(1- \beta_w \cos\theta)]\beta_w$ is
the apparent wave speed, and the observed wavelength can be greater or
less than the intrinsic wavelength depending on wave speed and viewing
angle $\theta$. In general, change in the observed wavelength is amplified
relative to change in the intrinsic wavelength with
\begin{equation}
 \lambda^{ob}(z_{ob}) = \lambda^{ob}(z\sin\theta) = { \sin\theta \over
1- \beta_w(z) \cos\theta}\lambda(z)~.
\end{equation}
In equation (3) $z$ is the intrinsic distance to the core where $z_{ob} \equiv z\sin\theta$ is the observed distance.  If the wave speed is
relatively low at low frequencies, i.e., $\gamma^2 \eta < 1$, then the
speed increases until it reaches the value given by equation (2) at
high frequencies.  In this case a helical wave advected along an
expanding jet can appear to increase in wavelength approximately
proportional to $R_j$ as the wave speed increases.  If $\gamma^2 \eta
>> 1$ there is a minimum in the wave speed near to the resonant
frequency and  there can still be significant change in the observed
wavelength.

Equation (1) can be rewritten to give the sound speed ratio in terms of the observed flow speed, wave speed ($\omega << \omega^*$) and viewing angle as
\begin{equation}
\eta \equiv (a_x/a_j)^2 = {[1 - (\beta_f^{ob} \sin\theta - \cos\theta)^2]\beta_{\omega}^{ob}
\over (\sin\theta + \beta_f^{ob} \cos\theta)(\beta_f^{ob} - \beta_{\omega}^{ob})
\sin\theta}~.
\end{equation}
Equation (2) can be rewritten to give the jet sound speed in terms of the
observed flow speed, wave speed ($\omega >> \omega^*$) and viewing angle as
\begin{equation}
\beta_s \equiv a_j/c = {\beta_f^{ob} -\beta_w^{ob} \over (\beta_f^{ob} + \beta_w^{ob})\cos\theta - (\beta_f^{ob} \beta_w^{ob} - 1)\sin\theta}~.
\end{equation}
Equation (3) can be rewritten to give the intrinsic wavelength in terms of the
observed wave speed, wavelength and viewing angle as
\begin{equation}
\lambda(z /\sin\theta) = \lambda(z_{ob}) = {1 \over
\sin\theta + \beta_w^{ob}(z_{ob}) \cos\theta} \lambda^{ob}(z_{ob})~.
\end{equation}
In equations (4 - 6) $\beta^{ob} \equiv v^{ob}/c$ is the apparent super
or subluminal flow or wave speed, and we have used $\beta =
[\beta^{ob}/(\sin\theta + \beta^{ob} \cos\theta)]$.  Provided
$\beta_f^{ob} \beta_w^{ob} > 1$, the superluminal case considered
here, a lower limit to the jet sound speed
\begin{equation}
\beta_s > {\beta_f^{ob} - \beta_w^{ob} \over \beta_f^{ob} + \beta_w^{ob}}
\end{equation}
can be found from equation (5) for $\sin\theta  \rightarrow 0$.  As the
viewing angle increases the sound speed increases and in the limit
$\beta_s \rightarrow 1/\surd 3$ a maximum viewing angle
\begin{equation}
\theta_{max} \approx {\surd 3 + 1 \over \beta_f^{ob}} -{\surd 3 - 1 \over \beta_w^{ob}}
\end{equation}
is obtained from equation (5) where we have used $\cos\theta \sim 1$,
$\sin\theta \sim \theta$, and assumed $\beta_f^{ob} \beta_w^{ob} >> 1$.
It can be shown that the supersonic limit, $\beta_s/\beta_f \le 1/\sqrt
3$, guarantees that $\beta_w^{ob}/\beta_f^{ob} > (1 - \beta_s/\beta_f)$
and $\theta_{max} > 0$.  As $\beta_w^{ob} \rightarrow
\beta_f^{ob}$ the less restrictive $\theta_{max} \sim 2/\beta_f^{ob}$
is recovered from equation (8).
  
Given that the observed prominent ``helical'' component proper motions
in the inner 3C\,120 jet are well defined, the largest uncertainty in
the modeling process lies in the assumed flow speed which could be as
small as $\beta_f^{ob}\mid_{min} \sim 5$ from the observed proper
motions of $\sim$~2.3~\masr\ found between components $o$ and $h$ at 43~GHz
\citep{G01} or as large as $\beta_f^{ob}\mid_{max} \sim 7.5$ from the
largest observed proper motion of $\sim$~3.4~\masr\ at 5~GHz
\citep{W01}.  The section of the jet that we consider also clearly
contains regions of different component spacing and motion.  Thus, we
are faced with the possibility of flow acceleration and changing sound
speeds along the jet.  Initially we shall consider two cases.  In both
cases let us assume $\beta_w^{ob} (\omega << \omega^*) \sim 0.6$ but
otherwise $\beta_w^{ob} (\omega >> \omega^*) \sim 4$ (appropriate to
inside 10~mas) or $\beta_w^{ob} (\omega >> \omega^*) \sim 5.5$
(appropriate to outside 10~mas).  Here we are identifying the
subluminal motions of the innermost observed components in the 3C\,120
jet with a minimum helical wave speed. Equations (4) and (5) then provide
preliminary estimates of the sound speeds and Figure 5 shows how
$\beta_s \equiv a_j/c$ and $a_x/c = \eta^{1/2}\beta_s$ vary as a
function of viewing angle, $\theta$, and observed flow speed,
$\beta_f^{ob}$. 
\begin{figure}[h]
\vspace{6.4cm}
\includegraphics{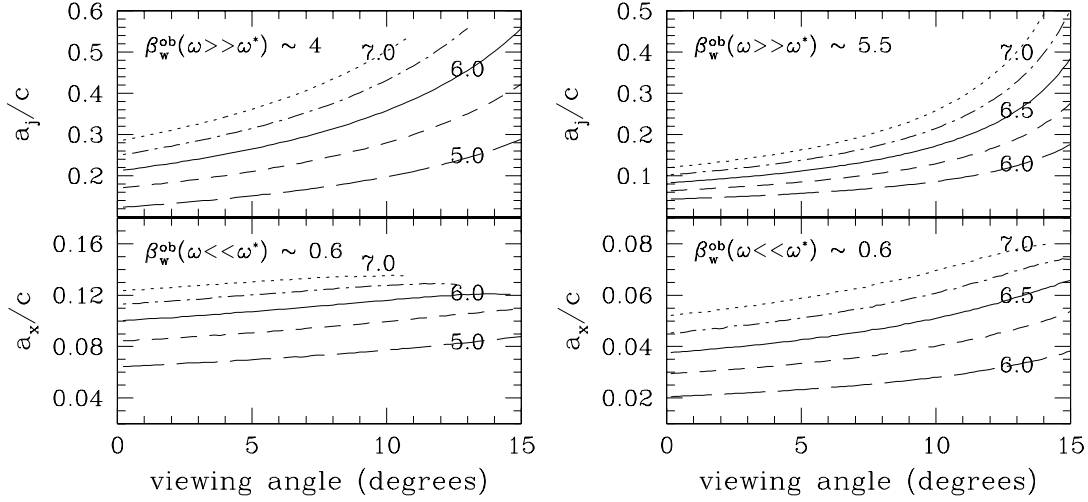}
\caption{\footnotesize \baselineskip 10pt Sound speeds from eqs.(4) \& (5) as a function of viewing
angle and apparent jet flow speed $\beta_f^{ob} = 5 - 7$. Jet sound
speed for two high frequency wave speeds, $\beta_w^{ob} (\omega >>
\omega^*) \sim$~4 (left top panel) \& 5.5 (right top panel).  The value
of the external sound speed that accompanies the high frequency wave
speed is shown in the panels immediately below.  Here in both cases we
assume that the observed low frequency wave speed is $\beta_w^{ob}
(\omega << \omega^*) \sim 0.6$.
\label{fig5}}
\vspace{-0.2cm}
\end{figure}
Examination of Figure 5 indicates that the external sound speed would
lie within the range $0.05 \le a_x/c \le 0.15$ and the internal sound
speed $a_j/c \ge 0.12$ for $\beta_w^{ob} (\omega >> \omega^*) \sim 4$.
Values are $0.02 \le a_x/c \le 0.08$ and $a_j/c \ge 0.04$ or factors 2
- 3 times lower for $\beta_w^{ob} (\omega >> \omega^*) \sim 5.5$.
Equations (7) and (8) with $\beta_w^{ob} \sim 4$ and $\beta_f^{ob} \sim
6$ suggest that the viewing angle is $\theta \le 15\arcdeg$ and we will
assume that $\theta = 15\arcdeg$ represents the largest possible
viewing angle.  The fact that implied sound speeds are lower for the
higher assumed wave speed beyond 10~mas suggests a jet cooled by
expansion.

When the observed spatial change in wave motion and wavelength imply a
helical wave with frequency within an order of magnitude of resonance,
the full wave dispersion relation must be used to fit the observed wave
motion and wavelength, and to estimate sound speeds.  In our initial
modeling we will follow the assumption that the jet can be modeled as
an isothermal constant speed expansion with isothermal external
medium.  For viewing angles $\theta = 15\arcdeg, 12\arcdeg, 9\arcdeg,
6\arcdeg$, and $3\arcdeg$\ we have considered two different cases that
bracket the likely flow and sound speed range.  In both cases we
consider the portion of the jet where the jet radius ranges from
$0.046$~mas~$\le R_j \le 1.56$~mas ($0.03$~pc~$\le R_j \le 1.01$~pc),
the observed distance ranges from $0.44$~mas~$\le z_{ob} \le
14.88$~mas, and the observed half-opening angle $\psi_{ob} = 6\arcdeg
\approx 0.105$~radian.

\noindent
For {\bf Case 1} we assume a flow speed $\beta_f^{ob} \approx 6$ with:
\begin{itemize}
\vspace{-0.2cm}
\item  The observed wave speed $\beta_w^{ob} \approx 0.7$ at $z_{ob} =
0.44$~mas and $\beta_w^{ob} \approx 4$ at $z_{ob} = 14.88$~mas.

\vspace{-0.2cm}
\item  At $z_{ob} = 14.88$~mas, the observed helical wavelength is
$\lambda^{ob} \sim 2.3R_j \sim 3.6$~mas.
\end{itemize}
\vspace{-0.2cm}
Here the observed wave speed ranges from typical of the proper motions
of the innermost components $s$ ($\beta^{ob} \sim 0.6$ @ $z_{ob} \sim
0.33$~mas) and $r$ ($\beta^{ob} \sim 0.9$ @ $z_{ob} \sim 0.55$~mas),
but does not achieve the proper motion of the ``helical'' components
($\beta_w^{ob} \sim 4$) for $z_{ob} \sim 2 - 10$~mas or the proper
motion of the ``helical'' components for $z_{ob} > 15$~mas
($\beta_w^{ob} \sim 6$). Additionally, the predicted wavelength for
$z_{ob} \sim 2 - 10$~mas is somewhat less than the observed
wavelength.  At least approximately this case is representative of the
jet close to the core.

\vspace{0.2cm}
\noindent
For {\bf Case 2} we assume a flow speed $\beta_f^{ob} \approx 6.5$ with:
\begin{itemize}
\vspace{-0.2cm}
\item  The observed wave speed $\beta_w^{ob} \approx 0.8$ at $z_{ob} =
0.44$~mas and $\beta_w^{ob} \approx 5$ at $z_{ob} = 14.88$~mas.

\vspace{-0.2cm}
\item  At $z_{ob} = 14.88$~mas, the observed helical wavelength is
$\lambda^{ob} \sim 3.1R_j \sim 4.3$~mas.
\end{itemize}
\vspace{-0.1cm}
Here the observed wave speed must begin somewhat above that of the
typical proper motions of the innermost components ($\beta_w^{ob} \sim
0.7$) in order to achieve the typical proper motion of the ``helical''
components ($\beta_w^{ob} \sim 4$) for $z_{ob} \sim 2 - 10$~mas, and to
nearly achieve the faster proper motion of ``helical'' components
($\beta_w^{ob} \sim 6$) for $z_{ob} > 15$~mas.  Here the predicted
wavelength is somewhat larger than the observed wavelength for $z_{ob}
\sim 2 - 10$~mas.

The above set of constraints along with the preliminary estimates of
allowed sound speeds, e.g., Figure 5, gets us into the correct region
of parameter space.  In general, we first obtain an estimate for the
jet sound speed by fitting the wavelength and wave speed at $z_{ob} =
14.88$~mas.  Here, where the wavelength is in the short wavelength and
high frequency regime, the wave speed is nearly independent of the
external sound speed and primarily depends on the jet speed and the jet
sound speed.  Typically this requires 10 or more trials at a single
viewing angle.  Note that we also obtain $\omega R_j/u$ at this
location and with $R_j$ and $u$ known we obtain the angular frequency
$\omega$.  Subsequently we find the external sound speed that would
yield the wave speed at $z_{ob} = 0.44$~mas and $\omega R_0/u$ given
our estimate of the jet sound speed.  Here the angular frequency,
$\omega$, of the wave is conserved but $\omega R_0/u$ is reduced
relative to $\omega R_j/u$ by the smaller jet radius.   A few
additional trials may be necessary as the value for the external sound
speed modifies slightly the required jet sound speed and value of
$\omega R_j/u$.  The resulting estimates for the sound speeds will be
accurate on the 10\% level, i.e., 10\% differences provide a similar
satisfactory fit.  Computing time is minimal, basically solving the
dispersion relation for different values of the sound speed, but the
non-linear nature of the dispersion relation near to resonance makes
this a procedure that would be difficult to automate.

Figure 6 shows a sample of the dispersion relation solutions that best
satisfy the set of conditions above along with the intrinsic and
observed wavelengths and wave speeds and Table 1 contains the associated
parameters.
\vspace{-0.3cm}
\begin{table}[h] 
 \begin{center}
 \caption{Jet Parameters for Case 1 \& Case 2}
 \vspace{0.2cm}
 \begin{tabular}{c c c c c c c c c c c c} \hline \hline
  $\theta$ & $\psi~(rad)$ & {\bf 1:} & $v_j/c$ & $\gamma$ & $a_j/c$ & $a_x/c$  
  & {\bf 2:} & $v_j/c$ & $\gamma$ & $a_j/c$ & $a_x/c$ \\  
 \hline
  15\arcdeg  & 0.0271 & | &  0.9910 &  7.47  &  0.460 &  0.095 & | &
  0.9943 &  9.37  &   0.400 &  0.040 \\
  12\arcdeg  & 0.0218 & | &  0.9874 &  6.31  &  0.330 &  0.110 & | &
  0.9900 &  7.09  &  0.250 &  0.045 \\
  9\arcdeg   & 0.0164 & | &  0.9864 &  6.09  &  0.260 &  0.130 & | &
  0.9885 &  6.61  &  0.180 &  0.050 \\
  6\arcdeg   & 0.0109 & | &  0.9882 &  6.53  &  0.210 &  0.150 & | &
  0.9895 &  6.92  &  0.130 &  0.055 \\
  3\arcdeg   & 0.0055 & | &  0.9927 &  8.29  &  0.150 &  0.170 & | &
  0.9935 &  8.78  &  0.090 &  0.065\\
\hline
  \end{tabular}
  \end{center}
\vspace{-0.3cm}
\end{table}

For {\bf Case 1} the observed wave speed, $\beta_w^{ob} \sim 0.7$,
corresponds to the wave speed at frequency $\omega R_0/v_j \sim 0.32$.
The implied angular frequency is $\omega \sim 1 \times
10^{-7}$~radian/sec for a periodicity of about 2 years.  The jet sound
speed is lower than would be suggested by eq.\ (5) in order to achieve
$\beta_w^{ob} \approx 4$ at $z_{ob} = 14.88$~mas where $\omega R_j/v_j
\sim 11$ as the jet expands by a factor $\approx 34$ over the modeled
region.  The frequency is above ``resonance'' and thus the value for
the external sound speed, found from exact solution to the dispersion
relation, varies from about 10\% less to 40\% more than is found using
eqs.\ (4) \& (5) as viewing angle decreases from 15\arcdeg\ to
6\arcdeg.  We note that the indicated external sound speed at the
3\arcdeg\ viewing angle would need to be $a_x/c \sim 0.155$ to give
$\beta_w^{ob}\mid_{min} \sim 0.6$.  The fact that we cannot obtain an
external sound speed estimate consistent with the required subluminal
motions at $\theta \lesssim 3$\arcdeg\ and the observed subluminal
acceleration would suggest that $\theta > 3$\arcdeg.  However, both observational errors and uncertainties in the interpretation of features
makes a firm lower limit to the viewing angle unlikely.
\begin{figure}[h]
\vspace{13.3cm}
\includegraphics{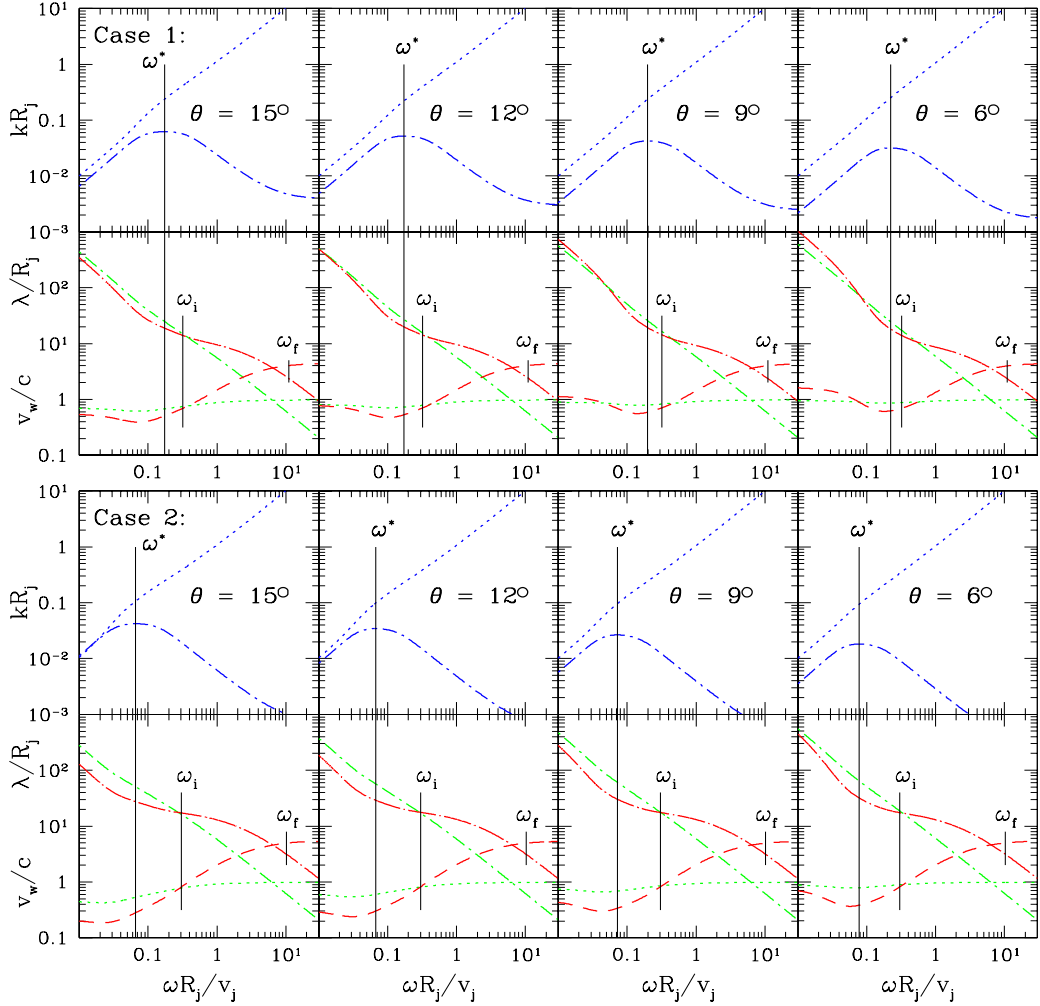}
\caption{\footnotesize \baselineskip 10pt Solutions $kR_j$ as a function of $\omega R_j/v_j$ for helical waves using parameters appropriate to viewing angles $\theta =$ 15\arcdeg, 12\arcdeg, 9\arcdeg\ \& 6\arcdeg\ for Case 1 (top panels) and Case 2 (bottom panels). \Blue{Dotted} (\Blue{Dash-dot}) lines indicate the real, $k_r$, (imaginary, $k_i$) part of the wavenumber.  The intrinsic (\Green{dotted}) and apparent (\Red{dashed}) wave speeds and intrinsic (\Green{short dash-dot}) and apparent (\Red{long dash-dot}) wavelengths are shown immediately below the appropriate dispersion relation solutions. The long, medium and short vertical lines mark $\omega^*$, $\omega_i (z_{ob} \sim 0.45$~mas), and $\omega_f (z_{ob} \sim 14.9$~mas), respectively.
\label{fig6}}
\vspace{-0.3cm}
\end{figure}

For {\bf Case 2}  the observed wave speed, $\beta_w^{ob} \sim 0.8$,
corresponds to the wave speed at frequency $\omega R_0/v_j \sim 0.30$.
The implied angular frequency remains $\omega \sim 1 \times
10^{-7}$~radian/sec for a periodicity of about 2 years.  The jet sound
speed is slightly lower than would be suggested by eq.\ (5) in order to
achieve $\beta_w^{ob} \approx 5$ at $z_{ob} = 14.88$~mas where $\omega
R_j/v_j \sim 10$.  The frequency is above ``resonance'' and and thus
the value for the external sound speed, found from exact solution to
the dispersion relation, varies from about 30\% less to 15\% more than
is found using eqs.\ (4) \& (5) as viewing angle decreases from
15\arcdeg\ to 6\arcdeg.  For this parameter set $\beta_w^{ob}\mid_{min}
< 0.6$ can be obtained at all viewing angles and no lower limit to the
viewing angle is implied by the observed subluminal acceleration.

In the above we have assumed that the subluminal motions of the
innermost 3C\,120 jet components would be representative of the helical
wave speed.  This would not be true if the helical wave was in the low
frequency limit at this location because low frequency propagation of
helical surface and pinch body modes can be very different.  However,
we find that the helical wave at this location is somewhat above
resonance.  A normal mode analysis shows that wave propagation at or
above the resonant frequency for pinch body modes and for helical twist
modes is approximately the same \citep{H00}.  Since our estimate is
based on a helical wave with initial frequency above resonance, there
will be only modest difference in the estimated external sound speed if
the subluminal component motions are modeled as pinch body mode
structures. The variation in external sound speed from this initial
position has little influence on our results. This is because the
modeled helical structure primarily lies in the high frequency regime
where wave speed is nearly independent of the external sound speed or
its variation.
  
\vspace{-0.7cm}
\subsection{Modeling Helically Twisted Fluid Flow}

As a helical twist is Kelvin-Helmholtz unstable, the displacement
amplitude of the jet surface grows according to
\begin{equation}
A = A_0~exp \left[ \int_{z_0}^{z}\ell(z)^{-1} dz \right]~,
\end{equation}
where $A_0$ is the displacement amplitude at $z_0$ and $\ell(z) \equiv
k_i^{-1}(z)$ is the spatial growth length.  Growth at the predicted
linear growth rates has been verified in the ``linear'' regime by
non-relativistic simulations \citep{SXH97,XHS00} and by
relativistic simulations \citep{P04}. 
The magnitude of accompanying
velocity and pressure fluctuations grows according to $A(z)/R_j(z)$,
and for constant jet expansion
\vspace{-0.1cm}
\begin{equation}
{A(z) \over R_j(z)} = {A_0 \over R_0}~{exp \left[ \int_{z_0}^{z}\ell(z)^{-1} dz \right] \over
\left[ 1 + (z - z_0)\psi/R_0 \right]}~,
\vspace{-0.1cm}
\end{equation}
where $R_j = R_0 + (z - z_0)\psi$.  For $\omega < \omega^*$ the growth
length is given by
\vspace{-0.1cm}
\begin{equation}
\eqnum{11a}
\ell(\omega) = \gamma \eta^{1/2}(\omega R_j/v_j)^{-1}R_j = \gamma M_j (\omega R_j/a_x)^{-1}R_j~.
\vspace{-0.1cm}
\end{equation}
At $\omega \sim \omega^*$ the minimum growth length
\vspace{-0.1cm}
\begin{equation}
\eqnum{11b}
\ell^* \gtrsim \gamma M_j R_j~
\vspace{-0.1cm}
\end{equation}
is somewhat larger than a straight extrapolation of the low frequency
result to the resonant frequency.  At frequencies $1< \omega/\omega^* < 10$ the growth
length rapidly increases with
\begin{equation}
\eqnum{11c}
\ell(\omega) \sim \ell^*(\omega/\omega^*)^m~,
\end{equation}
\setcounter{equation}{11}
where $m \lesssim 1$. At higher frequencies the growth rate can plateau
depending on jet and sound speeds.  The dependence of the growth length
on wave frequency relative to $\omega^*$ means that a maximum in
$A(z)/R_j(z)$ occurs on an expanding jet when $\omega > \omega^*$.
Note that the jet half opening angle, $\psi$, must be less than the
relativistic Mach angle, $\sim (\gamma M_j)^{-1}$, for
self-consistency.  Equation (10) reflects this self-consistency in that
$A(z)/R_j(z) \le A_0/R_0$ as $\psi \rightarrow (\gamma M_j)^{-1}$ and
no growth in pressure or velocity fluctuation is predicted.

Figure 7 shows how pressure and velocity fluctuations grow for a
viewing angle of 12\arcdeg.  The 1D cuts are made at locations $0.22
\le x/R_j \le 0.88$ and at $y = 0$.  Thus, $v_x$ is a radial velocity
and $v_y$ is a toroidal velocity component.
\begin{figure}[h!]
\vspace{13.5cm}
\includegraphics{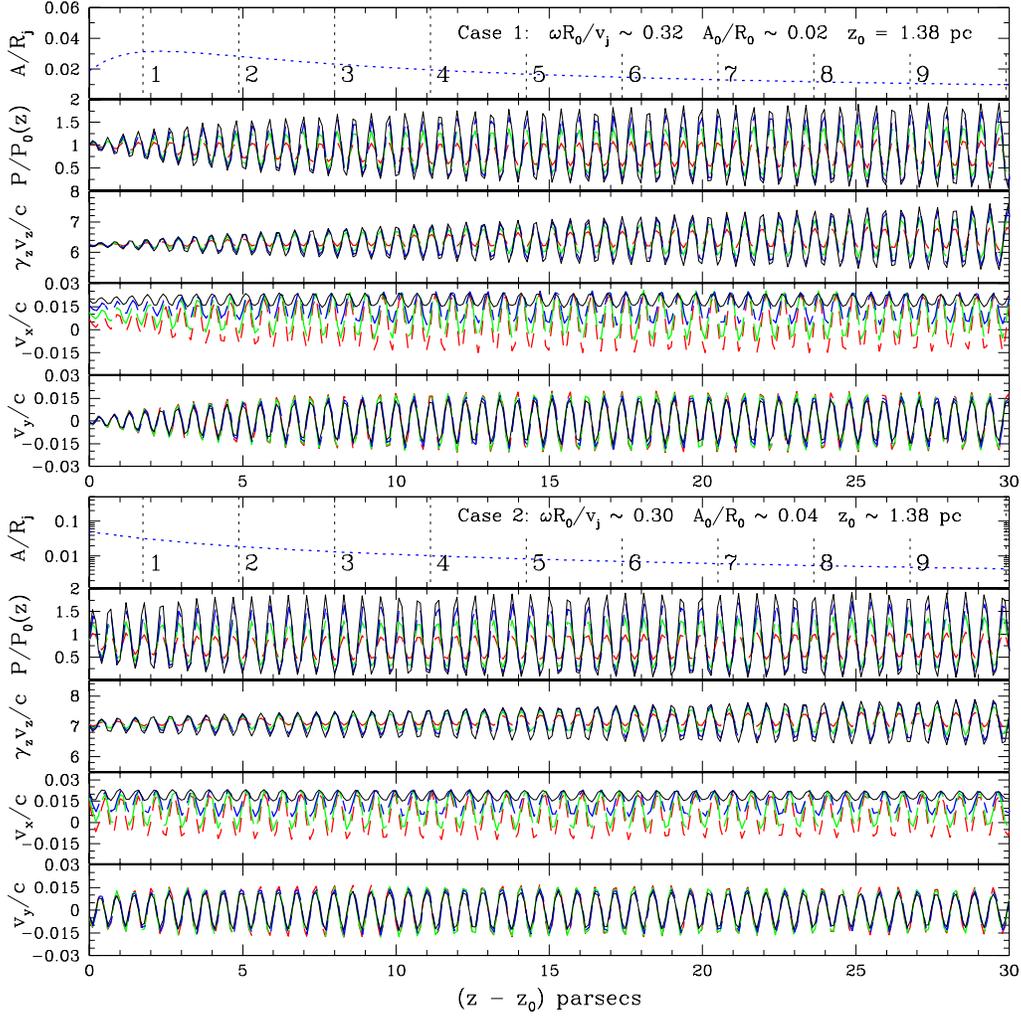}
\caption{\footnotesize \baselineskip 10pt The topmost panel of each set
shows the growth or decline of $A/R_{j}$ as a function of distance along
the jet axis for a viewing angle of $\theta = 12\arcdeg$. Three panels
below show the normalized pressure $P/P_0(z)$, and velocity components
$\gamma_z v_z/c$, $v_x/c$ and $v_y/c$. The vertical dotted lines mark
the projected separation from the core in mas.  1D cuts are at $x/R_j
=$ 0.22 (\Red{dash}), 0.44 (\Green{long dash}), 0.66 (\Blue{short-long dash}), 0.88 (solid).  
\label{fig7}}
\vspace{-0.4cm}
\end{figure}
Calculation of the pressure and velocity structure has used expressions
given in \citet{H00}. The spread in radial velocity from jet center
to jet surface indicates jet expansion and there is considerably more
fluctuation near jet center than at the jet surface.  Uniform $v_y$
across the jet indicates uniform sideways motion of the jet.  The axial
velocity shows more fluctuation at the jet surface than near the jet
center.  For all angles we have allowed the growth of perturbations
from some initial amplitude $A_0/R_0$ at position $z_0$ up to a
saturation amplitude assumed equal to the maximum allowed by the linear
analysis, i.e., $P \sim P_0(z) \pm P_0(z)$. For all viewing angles and
for both parameter sets we find that an initial amplitude can be chosen
such that saturation is achieved without imposing an arbitrary decline
in the growth rate at high frequency. Thus, we find that saturation can
occur naturally as a result of decline in the growth rate at high
frequency combined with jet expansion at constant opening angle.  For
Case 1 parameters where $\omega R_0/v_j \gtrsim \omega^*R_0/v_j$, $A/R_j$
increases before declining but for Case 2 parameters where
$\omega R_0/v_j >> \omega^*R_0/v_j$, $A/R_j$ declines.  Nevertheless,
the initial velocity and pressure perturbation at
$z_{ob} < 0.4$~mas that can lead to saturation is very small. In general,
for a given pressure fluctuation, velocity fluctuations are reduced for
lower values of the jet sound speed and higher values of the Lorentz
factor, i.e., higher relativistic Mach number $\gamma M_j$.  A saturation amplitude of $\Delta P \sim P_0$ corresponds to
flow with helical pitch at about half the relativistic Mach angle.  For
saturated helical waves with $\omega >> \omega^*$ displacement of the
jet surface $A << 0.1 R_j$.

\vspace{-0.7cm}
\subsection{Modeling Helical Intensity Morphology}
\vspace{-0.1cm}

In intensity modeling we do not attempt to reproduce details within the
jet or the exact positioning of ``helical'' components.  Line of sight
images are constructed assuming that a pseudo-synchrotron emissivity at
fixed frequency can be written as
\vspace{-0.2cm}
\begin{equation}
\epsilon_\nu \propto n_j^{1 - 2 \alpha}p_j^{2\alpha}(B~sin~\theta_B)^{1 +
\alpha}D^{2 + \alpha}
\vspace{-0.2cm}
\end{equation}
where $\theta_B$ is the angle of the magnetic field to the line of
sight and $D \equiv [\gamma(1 - \beta~cos~\theta)]^{-1}$ is the Doppler
boost factor for fluid flowing with speed $\beta$ at angle $\theta$ to
the line of sight.  The theoretically computed flow fields are used to
achieve correct inclusion of Doppler boosting effects on intensity
variations.  The magnetic field strength is assumed to decline and
fluctuate as $B \propto n_j^{2/3}$ appropriate to a disordered magnetic
field.  This form for the pseudo-synchrotron emissivity is one way to
calculate intensities when the synchrotron emitting particles are not
explicitly tracked \citep{CNB89}, and when radiative losses and
acceleration processes are not included.  \citet{JRE99} have shown that
pseudo-synchrotron intensities computed this way are not too different
from synchrotron intensities that include radiative losses and shock
acceleration processes, and so can provide an acceptable image of jet
structure.
\begin{figure}[h!]
\vspace{13.1cm}
\includegraphics{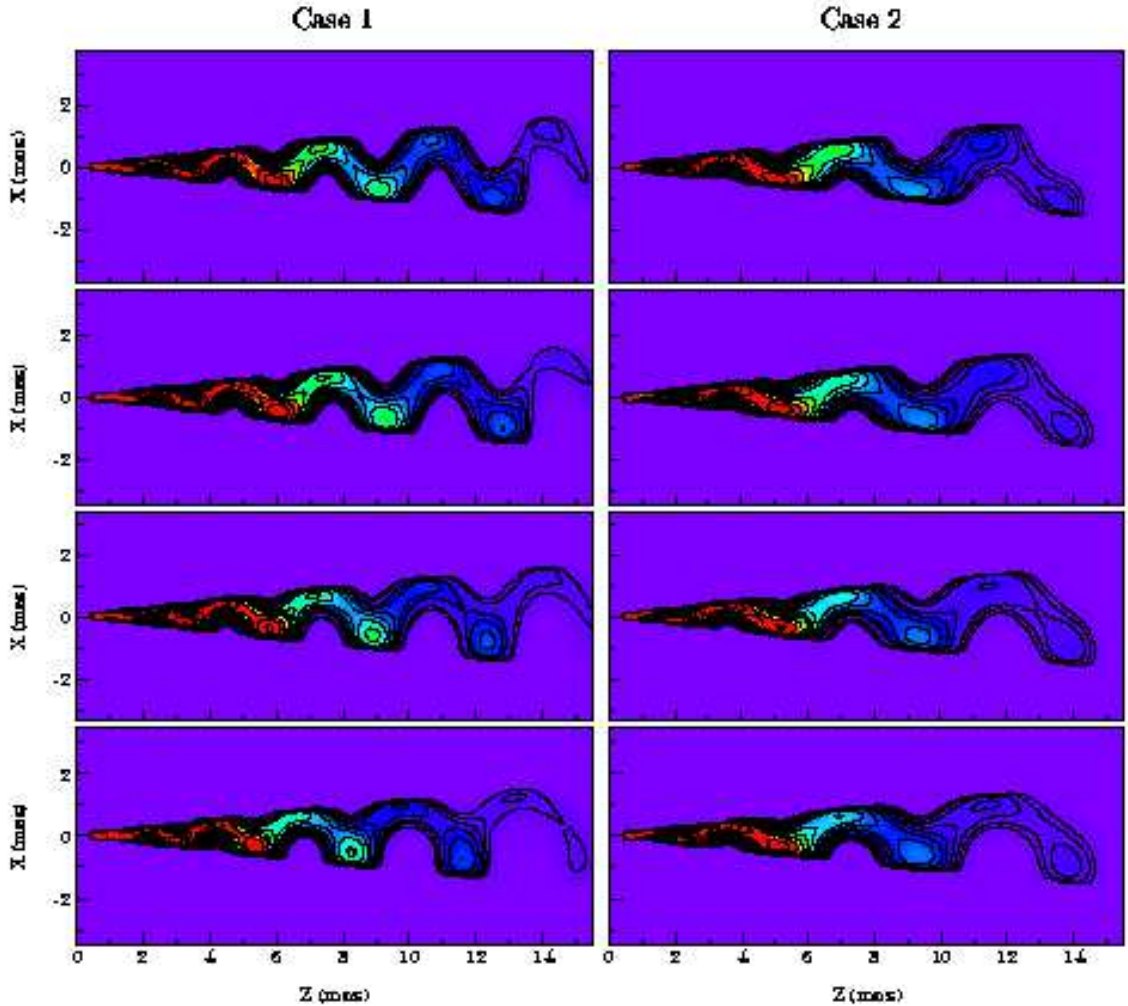}
\caption{\footnotesize \baselineskip 10pt Case 1 (left column) and Case 2 (right column) pseudo-synchrotron intensities for viewing angles 15\arcdeg, 12\arcdeg, 9\arcdeg\ and 6\arcdeg\ (top to bottom). Contours are in factors of $\sqrt 2$. Gray scale relates to intensity but gray scale and contours are adjusted relative to the maximum intensity
in each image and cannot be intercompared.
\label{fig8}}
\vspace{-0.3cm}
\end{figure}
In our intensity modeling we assume that the magnetic pressure is
negligible and the jet expands at fixed opening angle.  In what follows
we will model the jet assuming isothermal constant speed expansion,
i.e., conserve energy flux, and that the sound speed ratio, $a_x/a_j$
is constant.  Here the jet Mach number, $M_j \equiv v_j/a_j$, remains
constant and the external Mach number,  $M_x \equiv v_j/a_x$, also
remains constant with the jet in pressure balance with an external
cocoon medium.  Along the conical constant velocity isothermal jet the
particle number density and pressure decline $\propto R_j^{-2}$. For
the 3C\,120 jet we use $\alpha \sim 0.65$ where $I(\nu) \propto
\nu^{-\alpha}$.

Figure 8 shows the resulting pseudo-synchrotron intensity images for
four viewing angles.  Here the intrinsic jet becomes longer as the
viewing angle decreases so that the images represent the observed
projected jet.  All light travel time effects and Doppler boosting
effects are included.  The brightest radio emission traces the path of
the helically twisted high pressure ridge which effectively lies within
the conical jet's surface.
The helical pitch of the high pressure ridge is decoupled from the
helical pitch of the flow primarily as a result of rapid wave motion
when $\omega >> \omega^*$, and the intrinsic flow pitch angle is less
than would be inferred from the intrinsic helical pitch of the high
pressure ridge.  Line of sight effects mean that the brightest regions
will be on the south (lower) side of the jet where the high pressure
ridge approaches closer to the line of sight and increased somewhat by
increased Doppler boosting.  On the north (upper) side of the jet the
pressure ridge is stretched by projection and the brightness is
decreased somewhat because of reduced Doppler boosting.  The similarity
in appearance for the different viewing angles occurs because the flow
field and Doppler boost factor exhibit less angular variation at the
smaller viewing angles.  This reduced angular flow variation is a
result of the lower jet sound speed and higher Lorentz factor required
at the smaller viewing angles for the observed flow speed.

Helical component formation is enhanced at the shorter wavelengths of
Case 1 and is a combination of projection and enhanced Doppler
boosting.  However, our intensity variations which are the result of
projected adiabatic compressions show knot interknot variation less
than a factor of 4.  This is less than the variation indicated by the
5~GHz images (Figure 3), and considerably less variation than indicated
by the 22~GHz and 43~GHz images (Figure 1). This result suggests that
compressions resulting in the ``helical'' component formation in the
3C\,120 jet are not adiabatic. A secondary brightening on the northern
side of the jet between the brighter southern components is a
result of projection effects and relativistic aberration associated
with motion of the ribbon like high pressure region.  At the smaller
viewing angles the flow angle relative to the line of sight is
sufficiently increased at this location to reduce Doppler boosting and
reduce this secondary peak. The fact that the 22~GHz intensity
structure indicates a secondary brightening on the
northern side between helical components suggests that the 3C\,120 jet
lies at larger viewing angles.

\vspace{-0.7cm}
\section{Multiple Helical Frequencies and Adiabatic Expansion}
\vspace{-0.1cm}

The jump in component spacing seen in the 5~GHz images (Figure 4)
cannot be reproduced by a simple increase in observed wavelength of a
helical twist of single frequency.  We have investigated a combination
of harmonic frequencies, $\omega_2 \approx 2 \omega_1$, for the
isothermal jet expansion model and also for an adiabatic jet expansion
model in which flow acceleration occurs.  In what follows we restrict our
attention to the single viewing angle of 12\arcdeg.  

For a single frequency excited by precession of the central engine the
behavior of a helical twist is predictable over many orders of
magnitude of jet expansion.  Difficulties arise if multiple frequencies
are excited.  The largest uncertainty in dealing with multiple
frequencies lies in the interaction between multiple frequencies
associated with the same wave mode.  Numerical simulations have shown
that the faster growing higher order normal modes (elliptical,
triangular etc.) saturate, do not slow the growth of a helical mode
wave, and decline in amplitude as the longer wavelength helical twist
grows \citep{HCR97}.  Another simulation shows that pinch mode
waves show abrupt change to longer wavelength on a conically
expanding axisymmetric jet \citep{A01}.  Therefore we follow the
assumption that a high frequency saturated wave will decline in
amplitude as a lower frequency wave grows as specified by the
linear theory.

\vspace{-0.7cm}
\subsection{Isothermal Expansion Model}
\vspace{-0.1cm}

We use the isothermal jet parameters of Case 2 as we are interested in
structure beyond 10~mas and consider $\omega_1 R_0/v_j \sim 0.18$ and
$\omega_2 R_0/v_j \sim 0.36$ at $z_{ob} \sim 0.44$~mas.   This choice
of frequencies provides the closest approximation to the wavelengths
evident in 43, 22, and 5~GHz images (Figs.\ 1 \& 4).  
\begin{figure}[h!]
\vspace{13.1cm}
\includegraphics{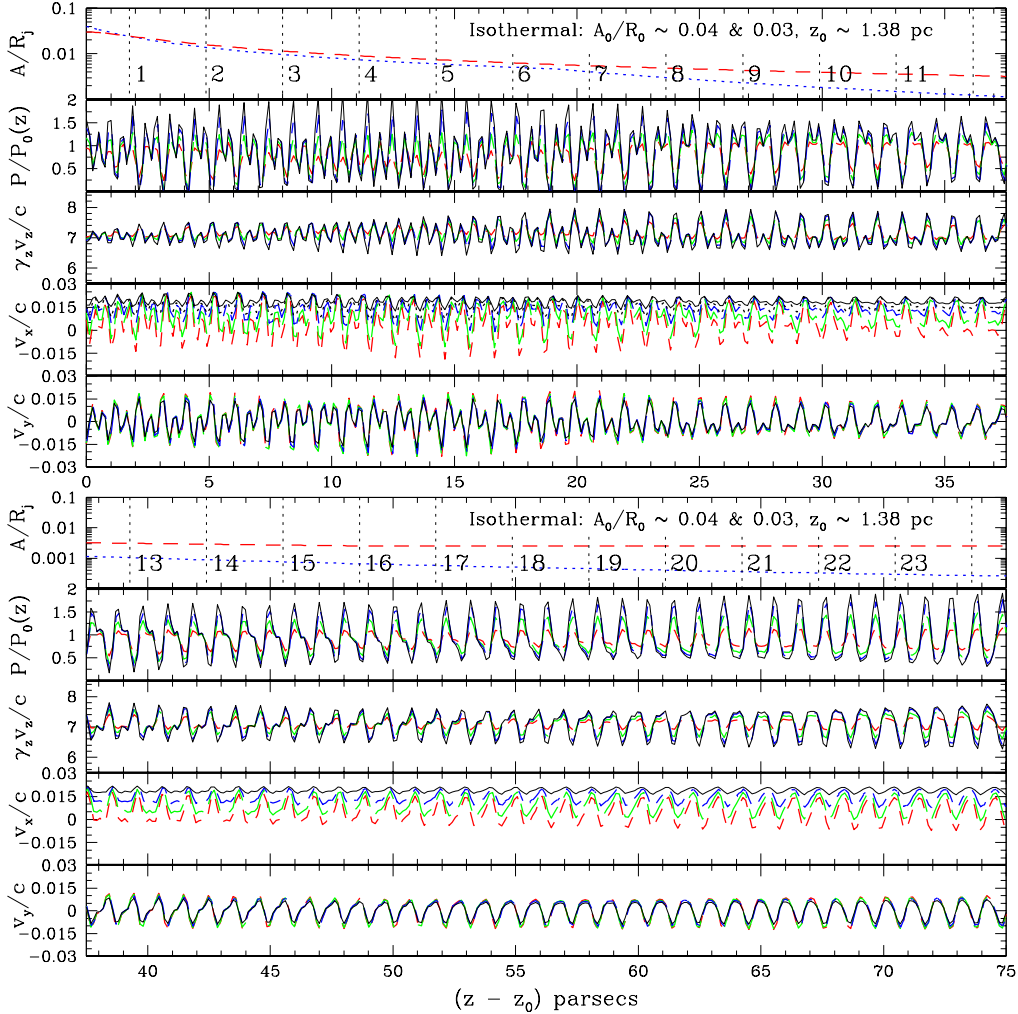}
\caption{\footnotesize \baselineskip 10pt The topmost panel of each set shows the growth or decline of
$A/R_{j}$ as a function of distance along the jet axis for a viewing
angle of $\theta = 12\arcdeg$. The \Blue{dotted} lines \& \Red{dashed}
lines show $A_2/R_j$ \& $A_1/R_j$, respectively. Three panels below
show the normalized pressure $P/P_0(z)$, and velocity components
$\gamma_z v_z/c$, $v_x/c$ and $v_y/c$. The vertical dotted lines mark
the projected separation from the core in mas.   1D cuts are at $x/R_j
=$ 0.22 (\Red{dash}), 0.44(\Green{long dash}), 0.66 (\Blue{short-long
dash}), 0.88 (solid).  
\label{fig9}}
\vspace{-0.2cm}
\end{figure}
The initial wave
amplitudes, $A_2/R_0 \sim 0.04$ and $A_1/R_0 \sim 0.03$, have been
chosen so that the lower frequency grows without constraint and as
specified by the linear theory.  The high and low frequency waves are
allowed to grow  at small core distances as specified by the linear
theory until the combined amplitudes are saturated.  At larger
distances the high frequency wave is damped as the low frequency wave
continues to grow as specified by the linear theory in order to keep
pressure fluctuations within the ``linear'' limit.  
This damping is ad hoc as the linear equations used here provide no
mechanism for damping.  The wave amplitudes along with the accompanying
pressure and velocity fluctuations are shown in Figure 9. In this
self-consistent calculation the initial low frequency amplitude
achieves pressure fluctuations of about 80\% of saturation at large
core distances.

A pseudo-synchrotron intensity image for the isothermally expanding jet
that corresponds to the fluctuations shown in Figure 9 is shown in
Figure 10.  This image shows modest helical components at the longer
wavelength, $\lambda^{ob}(\omega_1) \sim 8$~mas when $z_{ob} > 11$~mas,
but with the less than maximal short wavelength amplitude the modest
helical components that are evident when $z_{ob} < 8$~mas in Figure 8 do not
appear.  In part the lack of significant helical components is a result of
both this choice of viewing angle and the lower jet sound speed
required by Case 2, i.e., higher relativistic Mach number reduces
differential Doppler boosting effects.  The higher frequency wavelength
shows steady wavelength increase and $\lambda^{ob}(\omega_2) \sim 2 -
4$~mas when $z_{ob} \sim 1 - 11$~mas. 
It is clear that the basic
3C\,120 jet helical structure and motion can be reproduced in this
fashion and a larger or smaller viewing angle can recover modest
helical components in the inner 10~mas.  
\begin{figure}[h!]
\vspace{4.0cm}
\includegraphics{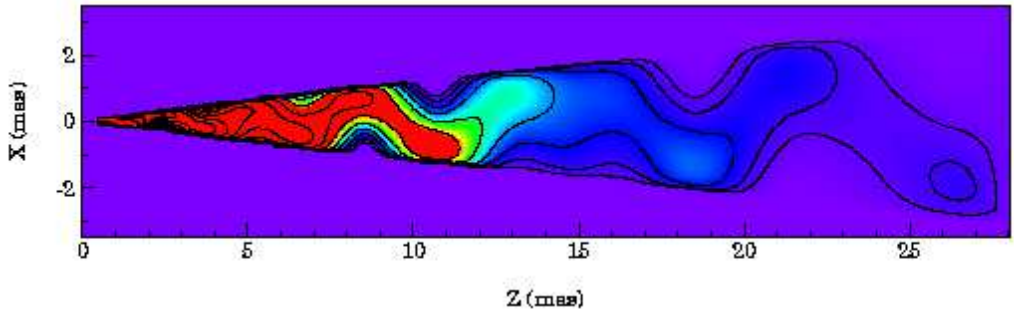}
\caption{\footnotesize \baselineskip 10pt Pseudo-synchrotron intensity for an isothermally expanding jet
with harmonic frequencies at viewing angle 12\arcdeg\ corresponding to
the fluctuations shown in Figure 9. Contours are in factors of 2.
\label{fig10}}
\vspace{-0.2cm}
\end{figure}
However, on the 3C\,120 jet the
higher frequency must saturate close to the core while still allowing
the lower frequency to saturate farther from the core as the higher
frequency is damped.  This requires that the higher frequency have a
larger initial growth rate than the lower frequency and this is not the
case for an isothermal expansion model with $\omega_2 > \omega_1 >
\omega^*$ and $k_i(\omega_2) < k_i(\omega_1)$ at $z_{ob} \ge
0.44$~mas.  Of course, the correct ordering of growth rates will be the
situation closer to the core.  Nevertheless, the high frequency wave
must be damped too close to the core and the low frequency wave does
not reach saturation for our best fit isothermal expansion model.

\vspace{-0.7cm}
\subsection{Adiabatic Expansion Model}
\vspace{-0.1cm}

The rapid initial growth in wave amplitude for the Case 1 isothermal
expansion model assumed to represent conditions nearer to the core and
the higher wave speeds for the Case 2 isothermal expansion model
assumed to represent conditions farther from the core suggests that a
combination of jet sound speed decline and jet acceleration could
reproduce the observed wavelength and proper motion behavior of helical
components along the inner 30~mas of the 3C\,120 jet.  Here we might
assume that the conditions associated with Case 1 are representative of
the innermost 2~mas of the observed jet.  To test this hypothesis we
have constructed a self-consistent adiabatic expansion model for a
viewing angle of 12\arcdeg\ with parameters varying as indicated in
Figure 11.
\newpage
\begin{figure}[h!]
\vspace{6.6cm}
\includegraphics{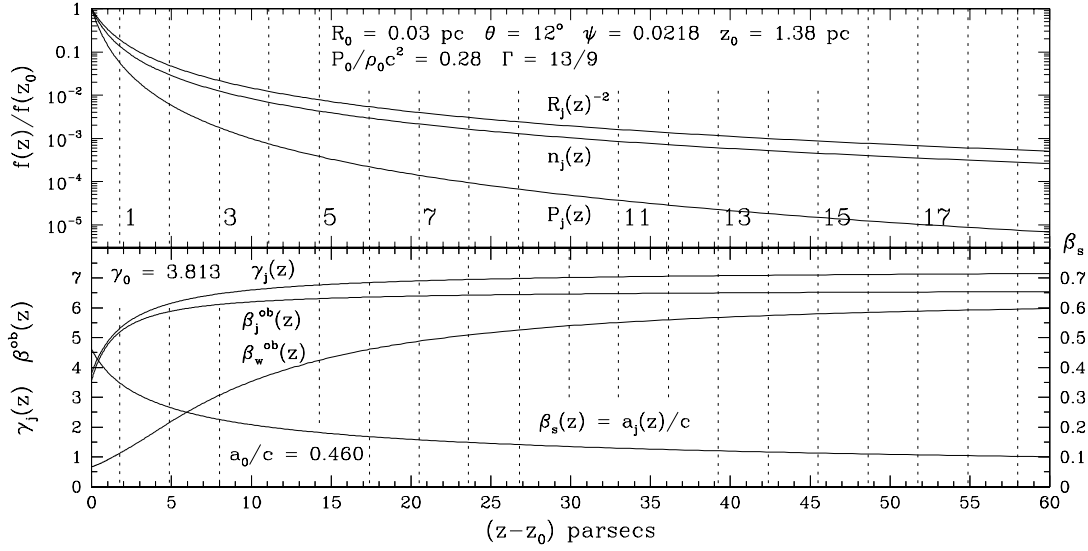}
\caption{\footnotesize \baselineskip 10pt The top panel shows the decline in jet density
$n_j(z)/n_j(z_0)$ and jet pressure $P_j(z)/P_j(z_0) \propto
n_j(z)^{13/9}$ along with $(R_0/R_j)^{2}$ for reference.  The bottom
panel shows the Lorentz factor, $\gamma(z)$, the jet sound speed,
$\beta_s = a_j(z)/c$, the observed flow speed, $\beta_j^{ob}(z)$, and
the observed helical component speed in the high frequency limit,
$\beta_{\omega >> \omega^*}^{ob}(z)$.  The vertical dotted lines mark
the projected separation from the core in mas.
\label{fig11}}
\vspace{-0.1cm}
\end{figure}

\noindent
To produce this model we have used the adiabatic wind equation
\vspace{-0.1cm}
\begin{equation}
(c^2/a^2 - 1)\frac{1}{\gamma}\frac{d\gamma}{dz} = \frac{2}{z}~,
\vspace{-0.2cm}
\end{equation}
along with the continuity equation
\begin{equation}
\gamma \beta n z^2 = \gamma_0 \beta_0 n_0 z_0^2~,
\end{equation}
and
\begin{equation}
\frac{a(z)}{c} = \left[ {{\Gamma P(z)}\over{\rho (z) c^2+ \frac{\Gamma}{\Gamma-1}P(z)}} \right]
^{1/2}~,
\end{equation}
where  $P(z) \propto n(z)^\Gamma$ with $\Gamma = 13/9$ appropriate to a
mixture of hot electrons and colder baryons \citep{S57}, and
$P_0/\rho_0 c^2 = 0.28$ to give $a_0/c = 0.460$ at $z_{ob} = 0.44$~mas
and with $v_0/c = 0.9650$. In the above, constant jet expansion at
fixed opening angle is assumed, i.e., $R_j \propto z$.
\begin{figure}[h!]
\vspace{5.70cm}
\includegraphics{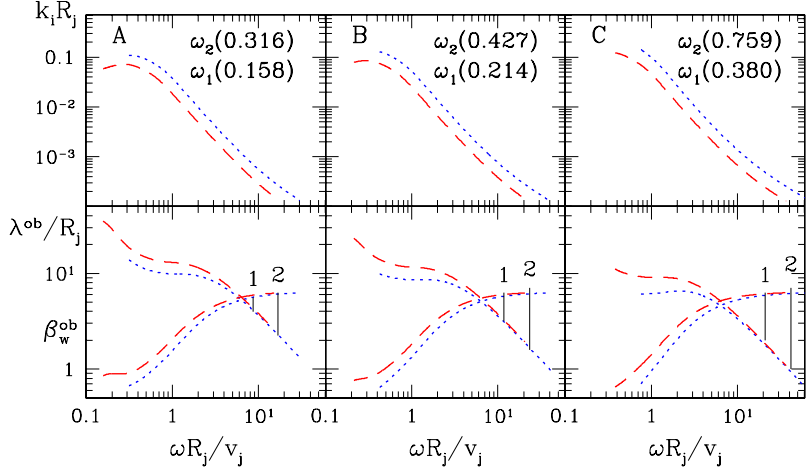}
\caption{\footnotesize \baselineskip 10pt The upper panels indicate the
growth rate, $k_i$, of harmonic frequency pairs beginning at $z_{ob} =
0.44$~mas.  The lower panels indicate the observed wavelength, $\lambda^{ob}$, and wave
speed, $\beta_w^{ob}$, with the vertical lines indicating where $z_{ob} \sim 25$~mas.
High frequency, $\omega_2$, (\Blue{dotted line}) and low frequency, $\omega_1$,
(\Red{dashed line}). \label{fig12}}
\vspace{-0.3cm}
\end{figure}

Along the adiabatically expanding jet the growth and wavelength of
helical frequencies must be tracked explicitly as conditions in the jet
and in the external medium change. 
In calculations the sound
speed ratio $a_x/a_j = 0.75$, which gives $\beta_w^{ob} \sim 0.7$ at
$z_{ob} = 0.44$~mas, will be assumed to remain fixed as $a_j$
declines.  There is no particular physical reason for this
choice but results are only weakly dependent on this assumption.
An
illustration of the growth rate, observed wave speed and observed
wavelength for three different initial harmonic frequency pairs,
$\omega_2 \approx 2 \omega_1$ is shown in Figure 12.  
All cases begin
at $z_{ob} = 0.44$~mas ($z_0 \sim 1.4$~pc) where $R_0 \sim 0.03$~pc and
$v_0/c = 0.9650$.  The position $z_{ob} \sim 25$~mas ($z \sim 80$~pc)
is indicated by vertical lines in the figure and corresponds to $\omega
R_j/v_j \sim 55 \omega R_0/v_0$.  The three panels span the possible
frequency range that will reproduce the observed wave speed and
wavelength behavior between the initial and final position.
Case C begins with the high frequency $\omega_2 = \omega^*$ at  $z_{ob}
= 0.44$~mas and is approximately the highest frequency for which
$k_i(\omega_2) > k_i(\omega_1)$.  In particular, case B with $\omega_1
R_0/v_0 = 0.214$ and $\omega_2 R_0/v_0 = 0.427$ provides the best
overall fit to the observed wavelengths and wave speeds.  On the
adiabatically expanding jet the high initial sound speeds and
relatively low initial Lorentz factor put $\omega_2 \sim \omega^*$
initially and $k_i(\omega_2) > k_i(\omega_1)$ for $z_{ob} \ge 0.44$~mas.

Wave growth and damping for case B along with the accompanying pressure
and velocity fluctuations is shown in Figure 13.  The higher frequency
with initial amplitude $A_2/R_0 \sim 0.0035$ is damped as the more
slowly growing low frequency wave is allowed to grow freely from its
initial amplitude, $A_1/R_0 \sim 0.0013$.  Amplitude growth here should
be compared to that shown in Figure 7 for high frequency growth near to
the core and isothermal expansion, and, in particular, to Figure 9 for
the development of a harmonic frequency pair on an isothermally
expanding jet.  Here initial amplitudes can be very small as the
initial growth rates are high and $\omega_2$ reaches saturation
quickly. Amplitude growth is consistent with high frequency growth to
saturation subsequently supplanted at larger distance by a more slowly
growing lower frequency.  Reduced velocity fluctuation at larger core
distance compared to the isothermal model for maximal pressure
fluctuation is a result of high Lorentz factor and low jet sound speed,
i.e., high relativistic Mach number $\gamma M_j$.  Overall, this
dynamical result is more consistent with the required behavior of rapid
high frequency growth followed by subsequent damping as the lower
frequency grows to saturation.

A pseudo-synchrotron intensity image for the adiabatically expanding
jet is shown in Figure 14.  The apparent wave speed and wavelength
undergoes a much more rapid initial increase than the isothermal Case 2
(Figure 8) from wavelength $\lambda^{ob} < 1$~mas at $z_{ob} < 2$~mas
to $\lambda^{ob} \sim 2.5 - 3.8$~mas when $z_{ob} \sim 2 - 8$~mas. No
additional wavelength or wave speed change for the higher frequency
wave occurs beyond about 5~mas from the core. The low frequency wave
dominates with $\lambda^{ob}(\omega_1) \sim 8$~mas when $z_{ob} >
14$~mas.  The apparent wavelength and wave speed of the adiabatic model
fits the observed wavelength and motions in the inner 30~mas of the
3C\,120 jet somewhat better than the isothermal model.
However, here the intensity falls much more rapidly than is observed on
the 3C\,120 jet.  While neither model produces significant helical
components at $z_{ob} < 10$~mas, the adiabatic jet shows no evidence
for even modest southern helical components at $z_{ob} > 15$~mas even
though the viewing angle is identical to the isothermal intensity image
in Figure 10.  The lack of helical components results from projection
effects associated with relativistic aberration and a reduction in
differential Doppler boosting.
\begin{figure}[h!]
\vspace{13.0cm}
\includegraphics{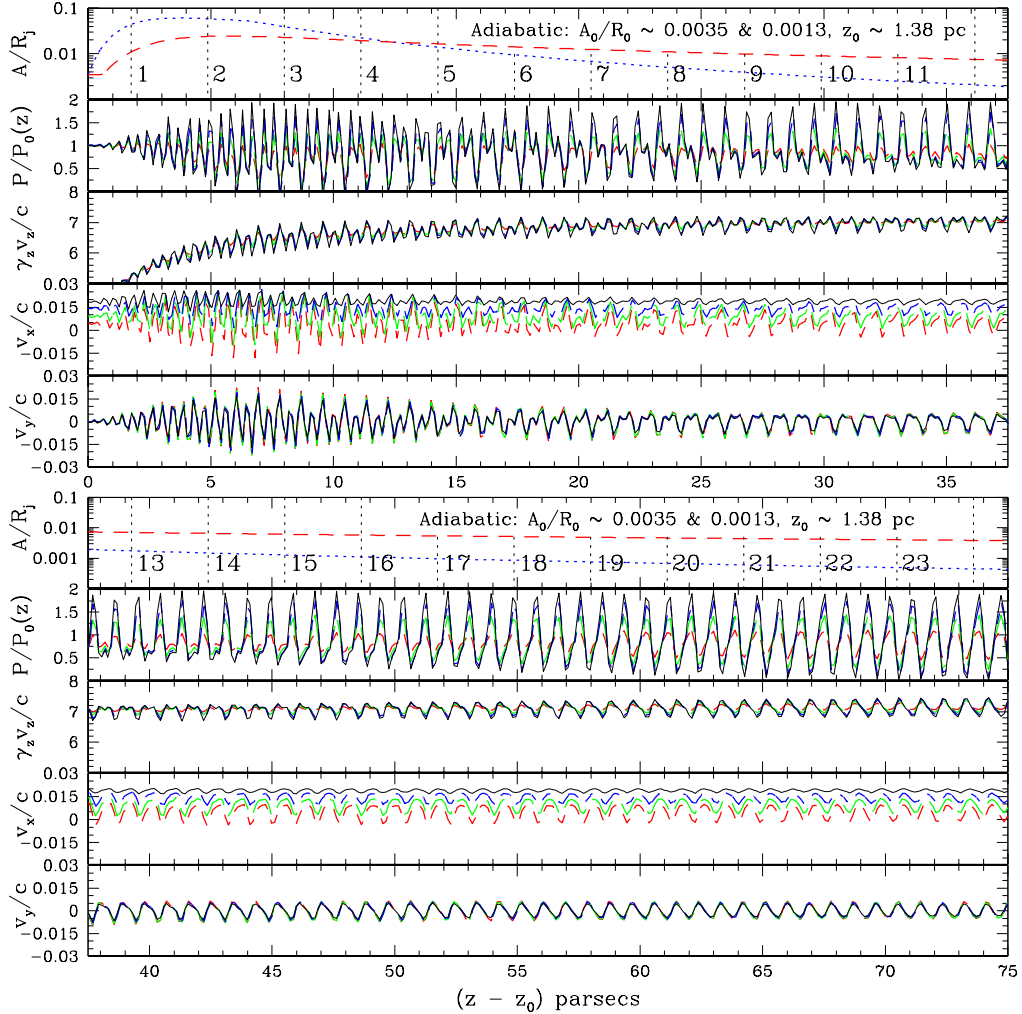}
\caption{\footnotesize \baselineskip 10pt The topmost panel of each set
shows the growth or decline of $A/R_{j}$ as a function of distance along
the jet axis for a viewing angle of $\theta = 12\arcdeg$.  The \Blue{dotted} lines \& \Red{dashed} lines show $A_2/R_j$ \& $A_1/R_j$, respectively. Three panels
below show the normalized pressure $P/P_0(z)$, and velocity components
$\gamma_z v_z/c$, $v_x/c$ and $v_y/c$. The vertical dotted lines mark
the projected separation from the core in mas.  1D cuts are at $x/R_j
=$ 0.22 (\Red{dash}), 0.44 (\Green{long dash}), 0.66 (\Blue{short-long dash}), 0.88 (solid).  \label{fig13}}
\vspace{-0.3cm}
\end{figure}
\begin{figure}[h!]
\vspace{4.1cm}
\includegraphics{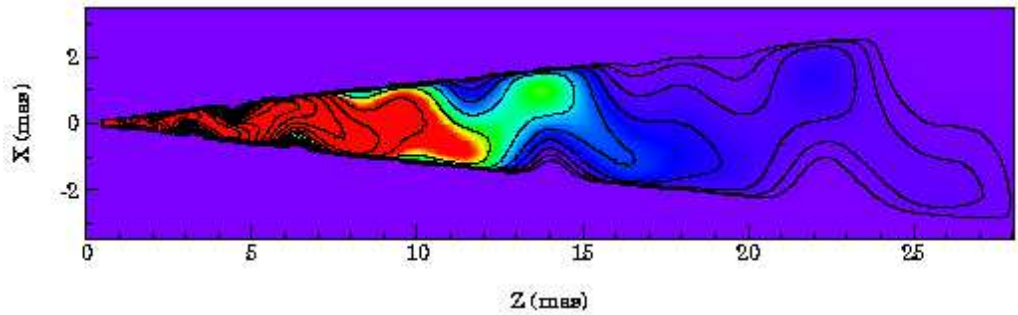}
\caption{\footnotesize \baselineskip 10pt Pseudo-synchrotron intensity for an
adiabatically expanding jet with harmonic frequencies at viewing angle 
12\arcdeg\ corresponding to the fluctuations shown in Figure 13. Contours are in factors of 2.
\label{fig14}}
\vspace{-0.2cm}
\end{figure}
Here the helical pattern is viewed at an
angle $\theta = 0.21$~rad $> 1/\beta_{\rm w}$ when $z_{ob} > 8$~mas (see
Figure 11). This has the effect of producing northern ``helical'' knots
at 14 and 22~mas. This implies that the jet is at a viewing angle less than 12\arcdeg. Differential Doppler boosting is reduced because the
relativistic Mach number rises and transverse velocity fluctuation
declines (see Figure 13).

\vspace{-0.5cm}
\subsection{An Optimized Isothermal Image}

Constraints on our modeling are relaxed somewhat if much less
acceleration in pattern speed is required, i.e., from about
$\beta_w^{ob} \lesssim 4$ when $z_{ob} \sim 2 - 8$~mas to about
$\beta_w^{ob} \sim 6$ when $z_{ob} > 14$~mas.  In this case the jet can
be closer to isothermal than our present results suggest, although some
sound speed decline and accompanying jet acceleration are still
required.  Such a result proves more consistent with the observed 5~GHz
intensity decline along the jet which is about a factor of 3000 (see
Figure 3). In order to make better comparison with the observed radio
intensities we have constructed a maximal pseudo-synchrotron intensity
image based on the isothermal Case 2 harmonic frequency pair. This theoretical image along with images
smoothed by Gaussian beams appropriate to 22 and 5~Ghz is shown in
Figure 15. 
\begin{figure}[h!]
\vspace{12.8cm}
\includegraphics{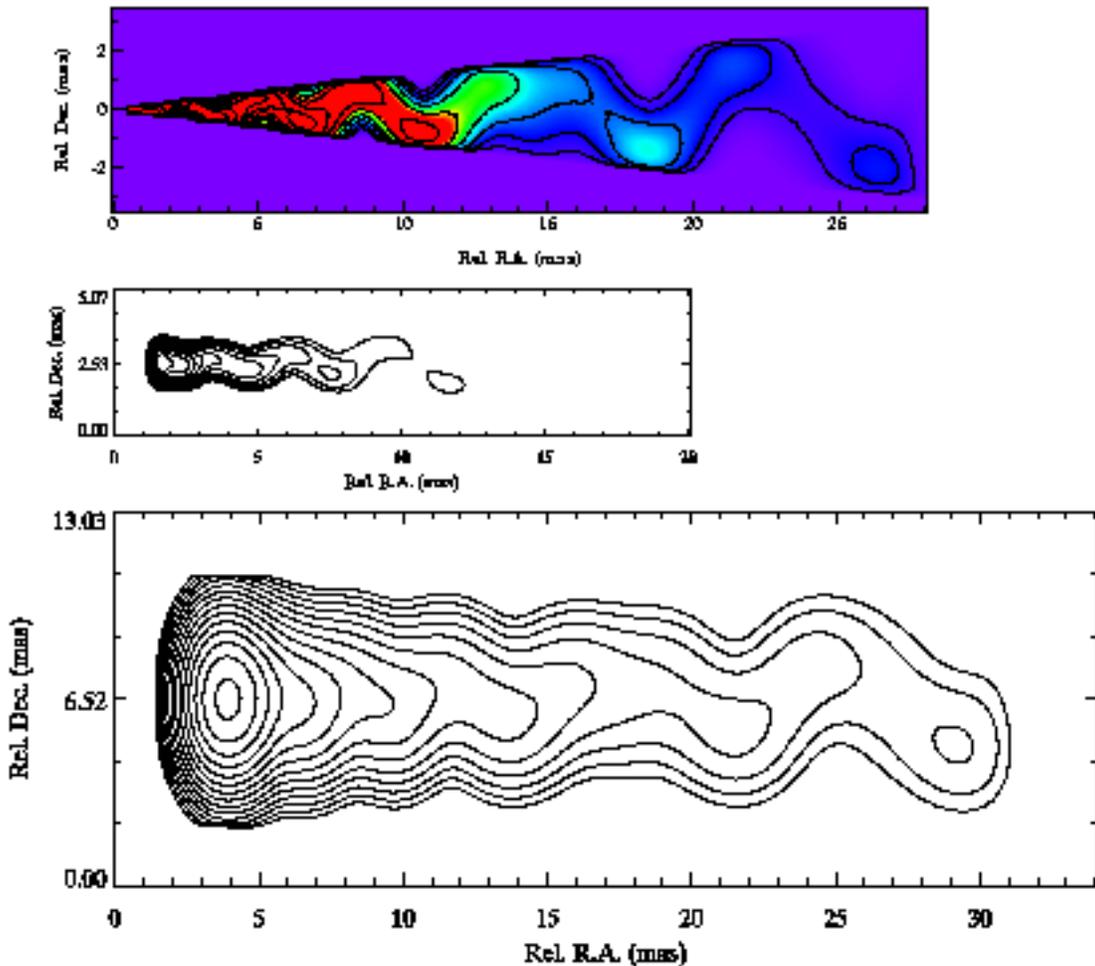}
\caption{\footnotesize \baselineskip 10pt Pseudo-synchrotron intensity images for an optimized pressure
fluctuation saturated isothermal expanding jet with harmonic
frequencies at viewing angle 12\arcdeg. The topmost is the initial
computed theoretical image. The middle image is smoothed by a $0.3
\times 0.6$~mas Gaussian beam appropriate to 22~GHz, has a peak of 8126
units/beam and the lowest contour is 0.01 units/beam.  The bottom is
image smoothed by a $1.3 \times 2.6$~mas Gaussian beam appropriate to
5~GHz, has a peak of 3949 units/beam, and the lowest contour is 0.1
units/beam.  In all images contours are in factors of 2.  The convolved images have considerably larger dynamic range than the observations.
\label{fig15}}
\vspace{-0.2cm}
\end{figure}
In
constructing the theoretical intensity image we have arbitrarily fixed
the amplitude of the lower frequency at $A_1/R_j = 0.0035$, i.e.,
achieves saturation at $z_{ob} > 20$~mas and allowed $A_2/R_j \le
A_2/R_0 \sim 0.04$ to decline while maintaining saturation of the
harmonic combination. When $z_{ob} < 7$~mas velocity and pressure
fluctuations are similar to the Case 2 viewing angle $\theta =
12\arcdeg$ image shown in Figure 8. This maximizes the fluctuations at
both small and larger scales.  

The 5~GHz image in Figure 15 shows a steeper intensity decline than
seen in the observations.  Some of this difference can be explained by
core absorption at 5~GHz not included in the Figure 15 image. Thus, the
typical intensity decline as indicated by images at 5~GHz is on the
order of isothermal.  This result when combined with the somewhat
better fit to overall dynamics and pattern speed provided by the
adiabatic model suggests decoupling between macroscopic heating of the
jet fluid as a whole, less than isothermal, and microscopic
energization of the radiating electrons, at least isothermal.  However,
it is clear that adiabatic compressions are insufficient to give the
observed helical components and the knot interknot variation observed
(particularly at the higher 22 and 43~GHz frequencies).  Different
viewing angle can result in somewhat higher knot interknot intensity
ratios at least in the innermost short wavelength region but not
enough. Note that only viewing angles less than 12\arcdeg\ are allowed
as larger angles are ruled out by relativistic aberration effects.
Thus, if helical components are entirely the result of projection and
Doppler boosting they are not adiabatic compressions.  The implication
is that the leading edge of the helical pattern is a shock
compression.  Note that in this case the intensity might scale
approximately with the jet density even if macroscopic heating is
insufficient to make the jet fluid isothermal.  We might also expect
the intensity to be more ribbon like than the existing adiabatic
compression as the pressure profile would be strongly peaked near to
the jet surface.  Furthermore, differential Doppler boosting effects
may be enhanced by a flow field more aligned with the shock face, i.e.,
helical flow pitch more closely coupled to the intrinsic pitch of the
helical shocked pressure and density enhancement.

\vspace{-0.7cm}
\section{Summary \& Discussion}

We have assumed that prominent components at $z_{ob} \sim 2 - 30$~mas
are associated with a projected helical twist.  These ``helical''
components indicate the wavelength and motion of the helical twist.
Nearly constant spacing, $\lambda^{ob} \sim 3$~mas, and similar motion,
$\beta_w^{ob} \sim 4$, of prominent components along with a ``saturated''
component intensity structure between $2 - 10$~mas suggests that the
motion of these components could be associated with a saturated helical
wave in the high frequency regime.  Between $10 - 15$~mas a
discontinuity in the regular pattern of components occurs and between
$15 - 30$~mas the component spacing indicates a potential helical
pattern with a twist wavelength $\lambda^{ob} \sim 8$~mas, or about
twice the wavelength inside 10~mas.  The increase in component spacing
is accompanied by an apparent increase in the motion of components to
$\beta_w^{ob} \sim 6$.   Non-helical components outline a curved path in
intensity contours between helical components and their motion is
assumed indicative of the flow speed $\beta_f^{ob} \sim 6 - 6.5$.

In general wave growth, wavelengths, and wave motions are consistent with
jet expansion that lies somewhere between adiabatic and isothermal but
closer to adiabatic.  Saturation can occur naturally on the expanding
jet as a consequence of jet expansion and declining growth rate of
helical waves at frequencies above resonance. Modeling the observed
component spacing and motion as a helical twist from 2 - 30~mas
requires two harmonic twist frequencies.  The higher frequency has a
period of $\lesssim 2$~years.  A typical intrinsic wavelength of the
superluminally moving helical twist produced by the two frequencies is
six times the jet radius.  Declining jet sound speed accompanied
by modest jet acceleration provides the appropriate conditions for
growth of multiple frequencies. Still it is necessary to assume a
damping mechanism for the high frequency wave that allows the low
frequency wave to grow to saturation.

Modeling of the helical twist at the $\theta = 12\arcdeg$ viewing angle
that has been extensively considered here, suggests a decline in jet
sound speed from $c/3 < a_j < c/\sqrt 3$ at $\sim 0.5$~mas from the
core to $0.1 c < a_j < 0.25c$ at $\sim 25$~mas from the core
accompanied by some acceleration in the jet flow from Lorentz factor
$\gamma \lesssim 5$ to $\gamma \gtrsim 7$.  At $z_{ob} \sim 15$~mas
$\beta_s \equiv a_j/c > 0.13$ is obtained in the limit $\sin\theta
\rightarrow 0$.  The initial sound speed estimate at $\sim 0.5$~mas
from the core increases to $a_j \sim c/\sqrt 3$ at a maximum possible
viewing angle of $\theta \sim 15\arcdeg$.  At the larger viewing angle,
higher typical Lorentz factor is required but additional jet
acceleration also occurs.  Thus, the initial Lorentz factor would not
be too much higher than for the 12\arcdeg\ viewing angle. However,
viewing angles  larger than 12\arcdeg\ can be ruled out.  At larger
viewing angles relativistic aberration produces northern instead of
southern ``helical'' knots when $\beta_{w}^{ob} > 5$ for $z_{ob} >
10$~mas. At smaller viewing angle the initial sound speed is less and
less flow acceleration would occur.  At a viewing angle of
6\arcdeg\ the jet sound speed might be about half of the estimates at
12\arcdeg, and the typical Lorentz factor would be comparable but with
less variation along the jet.  For $z_{ob} > 2$~mas the flow Lorentz
factor is in the range $\gamma \sim 6 - 8$ for viewing angles $\theta
\sim 4 - 14\arcdeg$.

Subluminal component motion at $z_{ob} < 1$~mas provides an estimate of
the sound speed in the cocoon medium outside the jet.  The subluminal
component motion requires an external (cocoon) sound speed $a_x > 0.1c$
at $z_{ob} \sim 0.5$~mas that is relatively independent of the viewing
angle. The value depends on the Lorentz factor at this location.  A
lower Lorentz factor requires a higher external sound speed but less
than the jet sound speed by a factor 1.5 - 5 depending on viewing
angle.  We cannot be more accurate as only the innermost $z_{ob} <
1$~mas of the jet provides any significant information on slow moving
pattern speeds.  This estimate was made by assuming that the innermost
subluminal components were associated with the helical twist.  However,
similar results will be obtained if the components are modeled as {\it
trailing} pinch structures triggered by passage of a shock down the jet
\citep{A01,G01}.  The combination of jet expansion, jet acceleration
and sound speed decline implied by our results automatically guarantees
that components associated with normal modes will show rapid
acceleration and increased spacing at $z_{ob} < 2$~mas.  More modest
increase in speeds and spacing occurs at larger core distance.

We cannot determine how the external cocoon sound speed might vary with
core distance from the observations considered here.  Basically any
determination of the change in external sound speed requires
information on low frequency or resonant wave motion far from the core
and here we see only high frequency waves (far above resonance) far
from the core.  Note that the present values for the external sound
speed and an assumed adiabatic decline are consistent with a hot slow
wind around a high speed hotter jet.  Such a picture may be reasonable
at the parsec to tens of parsec scales that we consider here.  A wind
with outflow speeds of a few tenths of lightspeed influences normal
mode dynamics and increases the transverse velocity fluctuation
produced by a saturated pressure fluctuation, e.g., \citet{HH03}.  This
effect would enhance differential Doppler boosting, possibly to
significant levels.

In earlier work \citet{W01} showed that a flow helical pitch comparable
to the intrinsic helical pitch implied by the observations can result
in too much knot interknot intensity variation as a result of
differential Doppler boosting.  In this case differential Doppler
boosting depended strongly on the viewing angle and thus knot interknot
intensity variation could be used to constrain the viewing angle.
Unfortunately, the present modeling of intensities as adiabatic
compressions does not deliver sufficient helical projection plus
differential Doppler boosting to produce the observed knot interknot
intensity variations at any allowed viewing angle.  Here our first
order accurate flow field helical pitch is strongly decoupled and much
less than the intrinsic helical pitch of the pressure and density
enhancement.  Differential Doppler boosting could be enhanced by a
significant wind around the jet.  Alternatively the apparent decoupling
between the macroscopic fluid properties implied by an accelerating
cooling jet and the microscopic radiating particle properties implied
by an isothermal decline in the intensity can be taken to indicate the
presence of a shock along the leading edge of the helix.  The presence
of a shock would more nearly align the flow field parallel to the shock
face and enhance differential Doppler effects.  The investigation of
these possible effects is beyond the scope of this paper.

Finally our dynamical modeling implies that macroscopic viscosity and
heating are relatively low along this jet, at least when compared to
the energization requirements of the radiating electrons implied by our
intensity modeling.  A low macroscopic viscosity is consistent with the
requirement that a relatively sharp velocity discontinuity between jet
and cocoon must be present in order for high frequency short intrinsic
wavelength helical patterns to exist. The requirement of sharp
velocity  discontinuity indicates a relatively low rate of mass
entrainment.  Broad velocity shear, indicative of higher macroscopic
viscosity and mass entrainment, would eliminate coherent structures
based on short normal mode wavelengths, e.g., \citet{B91}.  Note
however, that a shear layer increasing in thickness proportionally to
the jet radius may be one reason for a growing lower frequency wave
supplanting a higher frequency wave.  The higher frequency shorter
intrinsic wavelength would be damped as the jet expands and the shear
layer grows in thickness while the lower frequency longer intrinsic
wavelength could continue to grow.

\acknowledgements 
The European VLBI Network is a joint facility of European, Chinese,
South African and other radio astronomy institutes funded by their
national research councils.  P. Hardee would like to thank Jean Eilek
for numerous discussions concerning jet dynamics and acknowledges
partial support from the NASA/NSSTC to the University of Alabama.
J.L. G\'omez acknowledges support from the Spanish Ministerio de Educaci\'on
y Ciencia and the European Fund for Regional Development through grant
AYA2004-08067-C03-03.

\vspace{-0.5cm}
\appendix
\section{Observations at 5 GHz}

The 5 GHz images presented in Figure~2 are based on global VLBI
observations of 3C~120 made on 1999 February 11 (1999.11) and on 1999
September 21 (1999.72).  The data are from the ground stations
involved in Halca orbiting VLBI observations.  The data were recorded
in 2 baseband channels of 16 MHz each using an array, in February, of
19 antennas; the Very Long Baseline Array (VLBA --- 10 antennas), the
phased Very Large Array, the Green Bank 140 foot, and 7 antennas of
the European VLBI Network (Effelsberg, Jodrell Bank, Medicina, Noto,
Onsala, Westerbork, and Sheshan).  In September, the array was
somewhat smaller with Green Bank, Onsala, Noto, and Sheshan not
included.

Correlation was done on the VLBA correlator and data reduction was
done in AIPS.  The gain normalization during self calibration was
restricted to the VLBA antennas.  The flux density scale is good to
around 5\%.  A bandpass calibration was done to help achieve high
dynamic range by minimizing closure errors.  Many iterations of self
calibration and imaging were required because of the complexity of the
source --- much of the structure is outside the windows displayed in
this paper.  The final off-source rms noise levels were 49 and 41
$\mu$Jy beam$^{-1}$ for the February and September images respectively.
On-source errors, as usual, are expected to be significantly higher.

We thank John Benson of NRAO for his significant contributions to the
processing of these observations.  The eventual main publication of
these observations will show structures on significantly larger scales
than those displayed here.  They also will include the much higher
resolution images based on the Halca spacecraft data.


\vspace{-0.5cm}
\begin{thebibliography}{}
\baselineskip 12pt
\parskip 0pt

\bibitem[Agudo et al.(2001)]{A01} Agudo, I., G\'omez, J.L., Mart\'{\i},
J.M., Ib\'a\~nez, J.M., Marscher, A.P., Alberdi, A., Aloy, M.A., \&
Hardee, P.E. 2001, \apjl, 549, L183

\bibitem[Aloy et al.(1999b)]{A99b} Aloy, M.-A., Ib\'a\~nez, J.-M., Mart\'i, J.-M., G\'omez, J.-L., \& M\"uller, E. 1999, \apj, 523, L125

\bibitem[Aloy et al.(1999a)]{A99a} Aloy, M.-A., Ib\'a\~nez, J.-M.,
Mart\'i, J.-M., \& M\"uller, E. 1999a, \apjs, 122, 151

\bibitem[Aloy et al.(2000)]{A00} Aloy, M.-A., G\'omez, J.-L.,
Ib\'a\~nez, J.-M., Mart\'i, J.-M., \& M\"uller, E. 2000, \apj, 528, L85

\bibitem[Aloy et al.(2003)]{A03} Aloy, M., Mart\'{\i}, J., G\'omez, J.,
Agudo, I., M\"uller, E., \& Ib\'a\~nez, J. 2003, \apjl, 585, L109

\bibitem[Baldwin et al.(1980)]{B80} Baldwin, J.A., Carswell, R.F.,
Wampler, E.J., Smith, H.E., Burbidge, E.M., \& Boksenberg, A. 1980, 
\apj, 236, 388

\bibitem[Birkinshaw(1991)]{B91} Birkinshaw, M. 1991, \mnras, 252, 505

\bibitem[Clarke, Norman, \& Burns(1989)]{CNB89} Clarke, D.A., Norman, M.L., \& Burns, J.O. 1989, \apj, 342, 700

\bibitem[Duncan \& Hughes(1994)]{DH94} Duncan, G.C., \& Hughes, P.A.
1994, \apj, 436, L119

\bibitem[Falle \& Komissarov(1996)]{FK96} Falle, S.A.E.G., \& Komissarov, S.S. 1996, \mnras, 278, 586

\bibitem[G\'{o}mez et al.(1997)]{G97} G\'omez, J. L., Mart\'i,
J.-M., Marscher, A.P., Ib\'{a}\~{n}ez, J.-M., \& Alberdi, A. 1997, \apjl,
482, L33

\bibitem[G\'omez et al.(1998)]{G98} G\'omez, J.L., Marscher, A.P., 
Alberdi, A., Mart\'{\i}, J.M., \& Ib\'{a}\~{n}ez, J.M. 1998, 
\apj, 499, 221

\bibitem[G\'omez, Marscher, \& Alberdi(1999)]{G99} G\'omez, J.L.,
Marscher, A.P., \& Alberdi, A. 1999, \apjl, 521, L29

\bibitem[G\'omez et al.(2000)]{G00} G\'omez, J.L., Marscher, A.P., 
Alberdi, A., Jorstad, S.G., \& Garcia-Mir\'o, C. 2000, Science, 289,
2317

\bibitem[G\'omez et al.(2001)]{G01} G\'omez, J.L., Marscher, A.P., 
Alberdi, A., Jorstad, S.G., \& Agudo, I. 2001, \apj, 561, 161

\bibitem[Hardee(1987)]{H87} Hardee, P.E. 1987, \apj, 318, 78

\bibitem[Hardee(2000)]{H00} Hardee, P.E. 2000, \apj, 533, 176
 
\bibitem[Hardee(2003)]{H03} Hardee, P.E. 2003, \apj, 597, 798

\bibitem[Hardee, Clarke \& Rosen(1997)]{HCR97} Hardee, P.E., Clarke,
D.A., \& Rosen, A. 1997, \apj, 485, 533

\bibitem[Hardee \& Hughes(2003)]{HH03} Hardee, P.E., \& Hughes, P.A.
2003, \apj, 583, 116

\bibitem[Hardee et al.(2001)]{H01}Hardee, P.E., Hughes, P.A., Rosen,
A., \& Gomez, E. 2001, \apj, 555, 744

\bibitem[Hardee et al.(1998)]{H98} Hardee, P.E., Rosen, A., Hughes, P.A., \& Duncan, G.C. 1998, ApJ, 500, 599

\bibitem[Hughes et al.(2002)]{HMD02} Hughes, P.A., Miller, M.A., \& Duncan, G.C. 2002, \apj, 572, 713

\bibitem[Jones, Ryu, \& Engel(1999)]{JRE99} Jones, T.W., Ryu, D., \& Engel, A. 1999, \apj, 512, 105

\bibitem[Koide, Nishikawa, \& Mutel(1996)]{KNM96} Koide, S., Nishikawa, K.-I., \& Mutel, R.L. 1996, ApJ, 463, L71

\bibitem[Lobanov, Hardee, \& Eilek(2003)]{LHE03} Lobanov, A., Hardee,
P., \& Eilek, J. 2003, The Physics of Relativistic Jets in the CHANDRA
and XMM Era, New Science Reviews, 47, 629

\bibitem[Lobanov \& Zensus(2001)]{LZ01} Lobanov, A.P., \& Zensus, J.A. 2001, Science, 294, 128

\bibitem[Marscher et al.(2002)]{M02} Marscher, A.P., Jorstad, S.G.,
G\'omez, J.L., Aller, M.F., Ter\"asranta, H., Lister, M.L., \& Stirling,
A.M. 2002, \nat, 412, 625

\bibitem[Mart\'i, M\"uller, \& Ib\'a\~nez(1994)]{MMI94} Mart\'i, M\"uller, \& Ib\'a\~nez 1994, \aap, 281, L9

\bibitem[Perucho et al.(2004)]{P04} Perucho, M., Hanasz, M., Mart\'{\i}, J.M., \& Sol, H. 2004, \aap, in press

\bibitem[Steffen et al.(1995)]{S95} Steffen, W., Zensus, J.A.,
Krichbaum, T.P., Witzel, A., \& Qian, S.J. 1995, \aap, 302, 335

\bibitem[Stone, Xu, \& Hardee(2000)]{SXH97} Stone, J.M., Xu, J., \& Hardee, P.E.  1997, \apj, 483, 136

\bibitem[Synge(1957)]{S57} Synge, J.L. 1957, The Relativistic Gas, (Amsterdam:North-Holland)

\bibitem[Walker, Benson, \& Unwin(1987)]{W87} Walker, R.C., Benson,
J.B., \& Unwin, S.C. 1987, \apj, 316, 546

\bibitem[Walker et al.(2001)]{W01} Walker, R.C., Benson, J.B., Unwin,
S.C., Lystrup, M.B., Hunter, T.R., Pilbratt, G., \& Hardee, P.E. 2001,
\apj, 556, 756

\bibitem[Xu, Hardee, \& Stone(2000)]{XHS00} Xu, J., Hardee, P.E., \&
Stone, J.M. 2000, \apj, 543, 161

\end{thebibliography}
\end{document}